\documentclass[twocolumn,groupedaddress,showpacs,preprintnumbers,
amsmath,amssymb,aps,pre,floatfix]{revtex4-2}

\usepackage[dvipsnames]{xcolor}
\usepackage{layouts} 
\usepackage{listings}
\usepackage{dcolumn}
\usepackage{graphicx}
\usepackage{float} 

\usepackage[caption=false]{subfig} 

\begin{document}


\title{Understanding the swap Monte Carlo algorithm\\in a size-polydisperse model glassformer}


\author{Niklas Küchler}
\affiliation{Institut für Theoretische Physik II: Weiche Materie, 	Heinrich-Heine-Universität Düsseldorf,
	Universitätsstraße 1, 40225 Düsseldorf, Germany}

\author{Jürgen Horbach}
\affiliation{Institut für Theoretische Physik II: Weiche Materie, 	Heinrich-Heine-Universität Düsseldorf,
	Universitätsstraße 1, 40225 Düsseldorf, Germany}


\date{\today}

\begin{abstract}
The dynamics of a polydisperse model glassformer are investigated by augmenting molecular dynamics (MD) simulation with swap Monte Carlo (SMC). Three variants of the SMC algorithm are analyzed with regard to convergence and performance. We elucidate the microscopic mechanism responsible for the drastic speed-up of structural relaxation at low temperature. It manifests in a stepwise increase of the mean squared displacement when the time scale between the application of swap sweeps is significantly larger than a characteristic microscopic time scale. Compared to Newtonian dynamics, with the hybrid MD-SMC dynamics the glass transition shifts to a lower temperature and a different temperature dependence of the localization length is found. 
\end{abstract}


\maketitle

%
\section{Introduction}
The swap Monte Carlo (SMC) algorithm \cite{tsai1978structure, grigera2001fast, fernandez2007optimized} has been proven successful to efficiently equilibrate glassforming liquids with a size polydispersity \cite{gutierrez2015static, ninarello2017, fullerton2017density, ozawa2018random, brito2018theory, berthier2019bypassing, berthier2019efficient, berthier2019zero, klochko2020, lamp2022, gopinath2022diffusion, kuchler2022choice}. SMC introduces trial moves attempting to exchange the diameters between particles. In combination with conventional canonical Monte Carlo (MC) or molecular dynamics (MD) simulation, a dynamics is realized where the positional changes of the particles are accompanied by the fluctuations of their diameters. Ninarello \textit{et al.}~\cite{ninarello2017} optimized a model toward such a hybrid MC-SMC or MD-SMC dynamics to obtain ultrastable amorphous solid states. These states are comparable to those realized in experiments of structural glassformers and are far out of reach for any conventional MC or MD simulation.

Apart from the issue of generating well-equilibrated samples, the investigation of hybrid MC-/MD-SMC dynamics has been used to discuss fundamental aspects of the glass transition. SMC provides a proper sampling of the canonical ensemble on a frozen configuration (see also below) and thus its use does not affect thermodynamic properties of liquid and amorphous solid states in equilibrium. Based on this fact, Wyart and Cates \cite{wyart2017does} argued that the observed acceleration of the dynamics via SMC is not consistent with theories that explain the glass transition in terms of a growing static length scale. Via the analysis of the Hessian of soft- and hard-sphere systems, Brito {\it et al.}~\cite{brito2018theory} associated the speed-up due to SMC with the appearance of soft-elastic modes. They also demonstrated that the jamming transition is strongly altered by SMC. An interesting simulation study of a two-dimensional polydisperse soft-sphere system by Gopinath {\it et al.}~\cite{gopinath2022diffusion} introduced a swap model where only a selected fraction of particles can swap locally with neighboring particles. The authors interpreted the resulting ``defect diffusion'' in terms of a kinetically constrained lattice model. For such a lattice system, the kinetically constrained East model, Gutiérrez {\it et al.}~\cite{gutierrez2019accelerated} showed that swap moves lead to the suppression of dynamic heterogeneities, as also seen in polydisperse structural glassformers (see, e.g., Ref.~\cite{kuchler2022choice}).

The SMC dynamics has also been investigated in the framework of dynamic theories that predict a transition from a liquid to a non-ergodic amorphous solid, varying a control parameter such as temperature $T$ or packing density $\eta$.  In this manner, Szamel \cite{szamel2018theory, szamel2019theory} added a swap term to the mode coupling theory (MCT) \cite{gotze2009complex} equations for a binary hard-sphere system. Here, MCT predicts a liquid-solid transition at a critical packing fraction $\eta_c$ that depends on the size ratio of the hard-sphere species. While $\eta_c$ is around 0.515 without swaps, it increases up to about 0.535 with swaps. Note that a similar shift of the glass transition was found in the framework of a replica liquid theory by Ikeda {\it et al.}~\cite{ikeda2017mean, ikeda2019effect}. The dynamic MCT transition is intimately related to the cage effect, i.e., each particle is localized in a cage formed by the neighboring particles. According to MCT, in the non-ergodic solid state, the particles are trapped in their cages. However, upon decreasing the packing density toward $\eta_c$ (or increasing the temperature toward a critical $T_c$), the length $l$ that quantifies the localization of the particles in their cages increases toward a critical value $l_c$ at which the amorphous solid becomes unstable and transforms to a liquid state. In this sense, $l_c$ can be interpreted in terms of a Lindemann criterion for amorphous solids \cite{gotze2009complex}. Remarkably, Szamel's MCT as well as the replica approach by Ikeda {\it et al.}~indicate that with swaps the value of $l_c$ is significantly smaller than without swaps.

The latter theoretical approaches predict that swaps lead to a modification of the cage dynamics and a shift of the kinetic glass transition. However, these works do not provide a microscopic picture on how swaps change the structural relaxation of particles without affecting static equilibrium properties of the system. In the present work, we use hybrid MD-SMC simulations of a polydisperse model glassformer to address this issue and thus elucidate why swaps accelerate the dynamics so drastically.

A central idea of our work is to disentangle the effects of diameter fluctuations from the Newtonian dynamics of the particles. To this end, we first consider the application of swap moves on a given equilibrated configuration, keeping the positions of the particles fixed. We discuss different SMC algorithms, specified by the proposal probability with which the particle pairs for a diameter exchange are selected. Here, in particular, we find that a size-bias SMC scheme (or swap-sector scheme in Ref.~\cite{brumer2004numerical}), selecting only particle pairs that have a similar diameter, is more efficient than the standard SMC where one randomly chooses a pair of particles. With a diameter correlation function, we estimate the number of swap sweeps, $s_{\rm rel}$, required to thermalize the diameter configuration on a fixed set of particle positions (here, one sweep corresponds to $N$ trial swaps, where $N$ is the number of particles). At low temperatures, we find $s_{\rm rel} \approx 3$ for the size-bias SMC and $s_{\rm rel} \approx 13$ for the standard SMC. This information is used in a second step where we consider the full hybrid MD-SMC dynamics. First, we show that MD-SMC can be used to properly adjust the temperature of the system without the need for another thermostat. Then, to study the structural relaxation, we vary the time $t_{\rm MD}$ between swap sweeps. We identify the ``physically'' (however not computationally) most efficient MD-SMC scheme, for which $s_{\rm rel}$ sweeps are performed every integration time step $\Delta t$, i.e., $t_{\rm MD} = \Delta t$. Choosing a sufficiently large $t_{\rm MD}$ at low temperatures allows to infer the effect of the swap moves. After $s_{\rm rel}$ sweeps, the diameter permutation {\it instantaneously} imposes a new cage geometry around each particle. Then, during the subsequent MD part, the particles shift to new mean positions on a {\it microscopic time scale}. In this sense, this mechanism explains the drastic speed-up of the dynamics. It is reflected, e.g., in plateau steps of the mean squared displacement (MSD). The step-like behavior turns into a continuous increase of the MSD when $t_{\rm MD} = \Delta t$ is chosen. In the latter case we find, in agreement with the MCT prediction, that the glass transition shifts to a lower temperature in comparison to the pure Newtonian dynamics, accompanied by a smaller critical localization length $l_c$. 

The next section \ref{sec:model_and_simulation_details} is on the interaction model and the simulation details. Sec.~\ref{Sec:theory_swap_on_a_grid} presents the theory of SMC, assuming a fixed set of particle coordinates. Furthermore, we introduce three different SMC schemes with regard to the selection of particle pairs for the swap trial moves. In Sec.~\ref{Sec:numerical_results_swap_on_a_grid}, we study the relaxation dynamics of particle diameters using SMC on fixed particle configurations. The full hybrid MD-SMC dynamics is analyzed in Sec.~\ref{sec:MD_SMC}. Finally, we draw conclusions in Sec.~\ref{sec:conclusions}.

%
\section{Model and simulation details\label{sec:model_and_simulation_details}}
\subsection{Model
\label{sec:interaction_model}}

{\it Interaction model.} The model of the polydisperse glassforming system that we use in this work was proposed by Ninarello \textit{et al.}~\cite{ninarello2017}. We consider $N$ particles with varying diameters $\sigma = \sigma_1, \dots, \sigma_N$ and identical masses $m$ in a cubic box of volume $V=L^3$, using periodic boundary conditions. As specified below, the diameters are chosen according to a probability density $f$. In the following, positions and momenta of the particles are respectively denoted by the vectors ${\bf r}_i$ and ${\bf p}_i$, $i = 1, \dots, N$. The velocity ${\bf v}_i$ of particle $i$ is given by ${\bf v}_i = {\bf p}_i/m$. The particles move according to Hamilton's equations of motion with a Hamilton function $H=K+U$. Here, $K = \sum_{i=1}^N {\bf p}_i^2/m$ is the kinetic energy and the total potential energy $U$ can be written as
\begin{align}
U &= \sum_{i=1}^{N-1} \sum_{j>i}^N
u(r_{ij}/\sigma_{ij}) \,
\label{eq_potentialenergy},\\
u(x) &= u_0 \left(x^{-12} + c_0 + c_2
x^2 + c_4 x^4 \right)
\, \Theta(x_c - x) \,,
\label{eq:U_pair}
\end{align}
where the function $u$ describes the interaction between a particle pair $(i,\,j)$, separated by the distance $r_{ij} = |{\bf r}_i - {\bf r}_j|$. The argument of $u$ is scaled by the ``interaction diameter'' $\sigma_{ij}$ that is related to the diameters $\sigma_i$ and $\sigma_j$, as specified below.  With the Heaviside step function $\Theta$ a dimensionless cutoff $x_c = 1.25$ is introduced. The unit of energy is defined by $u_0$. The constants $c_0=-28 /x_c^{12}$, $c_2=48/x_c^{14}$, and $c_4=-21/ x_c^{16}$ ensure continuity of $u$ at $x_c$ up to the second derivative.

The interaction diameter $\sigma_{ij}$ introduces a \textit{non-additivity} of the particle diameters,
\begin{equation}
\sigma_{ij} = \frac{\sigma_i + \sigma_j}{2}
\left( 1 - 0.2 |\sigma_i - \sigma_j| \right) \,,
\label{eq:non_additivity}
\end{equation}
which is a significant ingredient to the model to suppress crystallization and demixing \cite{ninarello2017}. This is especially important when the swap Monte Carlo is used, since this algorithm provides the equilibration of samples at very low temperatures, where models with additive diameters become increasingly prone to crystallization.

\textit{Polydispersity.} Now we specify how we select the particle diameters $\sigma_i$.  The target distribution $f$ of the diameters is defined via the probability density
\begin{equation}
f(s) =
\begin{cases}
    A s^{-3}, & \sigma_{\rm m} \leq s \leq \sigma_{\rm M}, \\
    0, & \mathrm{otherwise}.
\end{cases}
\label{eq_fsigma}
\end{equation}
Here, a minimum $\sigma_{\rm m}$ and maximum diameter $\sigma_{\rm M}$ are introduced. The normalization condition $\int f(s)\,ds = 1$ sets $A = 2 / (\sigma_{\rm m}^{-2} - \sigma_{\rm M}^{-2})$. The unit of length is defined as the expectation value of the diameter, $\bar{\sigma} = \int \sigma f(\sigma)\,d\sigma$.  This implies $\sigma_{\rm M} = \sigma_{\rm m} /(2 \sigma_{\rm m} - 1)$. The distribution $f$ has one degree of freedom left, which is fixed by the choice $\sigma_{\rm m} = 0.725$. Then the upper bound is given by $\sigma_{\rm M} = 29/18 = 1.6\overline{1}$ and the amplitude by $A = 29/22 = 1.3\overline{18}$.  The degree of polydispersity $\delta$ can be defined via $\delta^2 = \int (s - \bar{\sigma})^2 f(s)\,ds/\bar{\sigma}^2$, so that $\delta \approx 22.93\%$. In our work the ratio $\sigma_{\rm M}/\sigma_{\rm m} = 20/9 = 2.\overline{2}$ deviates by less than 0.24\% from the value $2.219$ reported in Ref.~\cite{ninarello2017}.

\textit{Deterministic diameter choice.} In Ref.~\cite{ninarello2017} the diameters $\sigma$ were chosen randomly and independently from the density $f$. Different from this stochastic approach, we use a deterministic method that we extensively compared to the stochastic approach in Ref.~\cite{kuchler2022choice}. The deterministic method has the following advantages: (i) It leads to smaller finite-size effects, as the ``most representative sample''~\cite{santen2001liquid} is used for any system size $N$.  Statistical outliers are prevented in this way. This is especially important for glassforming liquids at low temperatures, which are very sensitive to density fluctuations. (ii) The histogram of the diameters converges to $f$ faster than for the stochastic approach.  (iii) A quenched disorder in the diameters is excluded, which would otherwise be present and superimpose sample-to-sample fluctuations. The latter point is not crucial to our analysis though, since we do not investigate sample-to-sample fluctuations here.

For the deterministic method the $N$ diameters are constructed as follows. In the first step, we introduce $N+1$ equidistant nodes $h_i = i/N$, $i=0, \dots, N$, along the codomain $[0,1]$ of the cumulative distribution function $F(s) = \int_{-\infty}^s f(\sigma)\,d\sigma$. The pre-images $s_i = F^{-1}(h_i)$ are well-defined, since $F$ is strictly monotonic and thus bijective when restricted to $[\sigma_m, \sigma_M]$. Finally, the diameters $\sigma_i$ are defined via $\sigma_i^3 = N \int_{s_{i-1}}^{s_i} \sigma^3 f(\sigma)\,d\sigma$. This scheme provides the same set of diameters for each sample. More details can be found in Ref.~\cite{kuchler2022choice}.

\subsection{Simulation details
\label{sec:equilibration_protocol}}
\textit{Simulation methods.} In all simulated samples, the number density is fixed to $N/V = 1$. The temperature $T$ is used as a control parameter. For the parts that include molecular dynamics (MD) simulations, we numerically integrate the equations of motion via the velocity form of the Verlet algorithm with a time step $\Delta t = 0.01\,t_0$. Here, $t_0 = \bar{\sigma} \sqrt{m/u_0}$ defines the unit of time.  Eventually, for the hybrid scheme (MD-SMC) combining MD with SMC as introduced in \cite{berthier2019efficient}, we apply $N \times s$ elementary SMC trials after every $t_\mathrm{MD}$ simulation time of MD dynamics. Here, $N$ trials define one sweep, so that $s$ defines an SMC density in a system-size-independent way. One elementary SMC trial refers to an attempt to exchange the diameters of a single pair of particles according to the Metropolis criterion. Which of the $N (N-1)/2$ pairs are chosen depends on the \textit{proposal probability} (also called \textit{a priori probability}), defining the specific SMC variant. Three different variants will be discussed in Sec.~\ref{Sec:theory_swap_on_a_grid}. Unless noted otherwise, we apply the standard SMC variant for which a random particle pair is chosen.

\textit{Equilibration protocol.} To equilibrate the samples, we use the hybrid MD-SMC scheme with $t_\mathrm{MD} = 0.25$ and $s = 1$. \textit{Only} during equilibration, we couple the system to the Lowe-Andersen thermostat \cite{Lowe_Andersen_T_different_masses} for identical masses $m$ with a frequency $\Gamma_\mathrm{T} = 4$ and a cutoff $R_\mathrm{T} = x_c$. For different system sizes (specified by the number of particles, $N = 256$, $500$, $2048$, and $8000$), we prepare $60$ samples, respectively. Each sample is equilibrated at different temperatures $T$ according to the following protocol. We start with the assignment of $N$ diameters as described in the previous subsection. The particles are placed on a face-centered-cubic lattice in the cubic box of length $L$, eventually with cavities in the case that $N \neq 4 n^3$ for all natural numbers $n$. The velocities ${\bf v}_i$ are initialized with a constant absolute value as ${\bf v}_i^2 = 3 T/m$ at the very high temperature $T=5$, but each with a random orientation. We subtract the mean momentum $\sum_{i=1}^N {\bf p}_i/N$ from each ${\bf p}_i$ to set the total momentum vector to zero. Then, the initial crystal is melted for a simulation time of $t = 2 \times 10^3$ with a short time step $\Delta t = 10^{-3}$ while the hybrid MD-SMC scheme and the Lowe-Andersen thermostat is applied. After that, we cool to $T=0.3$ for the same duration. Then, for each of the target temperatures $T$, a long run for a time span $t=10^5$ is performed with a time step $\Delta t = 0.01$. Then, we switch off the SMC algorithm and continue each simulation for another $t=0.8 \times 10^5$, during which the mean total energy $\bar{H}$ and fluctuations with the standard deviation $\mathrm{std}(H)$ are calculated for each run separately. After that, we simulate for $t=0.2 \times 10^5$. Here, as soon as the condition $|H-\bar{H}|< 0.01 \times \mathrm{std}(H)$ is satisfied, we switch off the thermostat to ensure that the energy $H$ is constant for the remaining time.

As shown in Ref.~\cite{kuchler2022choice}, for temperatures $T \gtrsim 0.06 \equiv T_\mathrm{g}^\mathrm{SMC}$, the prepared samples are fully equilibrated. Here, $T_\mathrm{g}^\mathrm{SMC}$ is the glass-transition temperature for MD-SMC. We do not observe any signs for ordering processes above $T_\mathrm{g}$, consistent with Ref.~\cite{ninarello2017}. As a reference point, for pure MD simulations in the microcanonical ensemble ($NVE$), the glass-transition temperature is around $T \approx 0.10$.

We draw pseudorandom numbers with the \textit{Mersenne Twister} algorithm~\cite{matsumoto1998mersenne}. A different seed for each sample is used to ensure independent sequences. 


\section{SMC on a frozen configuration \label{Sec:theory_swap_on_a_grid}}

In this section, we give an explicit mathematical description of the swap Monte Carlo algorithm (SMC) when applied to a fixed set of coordinates. By exchanging particles, SMC samples from a constrained phase space, the space of all particle permutations of a given configuration. Mathematically, SMC represents a discrete-time Markov chain, created via the Metropolis-Hastings algorithm \cite{robert1999monte} with the canonical distribution as a target distribution. We analyze three different SMC variants that only differ with respect to the {\it proposal probability}, i.e.~the selection of particle pairs: (i) the standard SMC in Sec.~\ref{sec:ordinary_SMC}, which allows transpositions between \textit{all particles}, (ii) a local SMC, for which only \textit{neighboring particles} are exchanged in Sec.~\ref{sec:local_SMC}, and (iii) a size-bias SMC, which only selects particles with \textit{similar diameters} in Sec.~\ref{sec:size-bias_SMC}. We discuss under which conditions each SMC variant converges.

\subsection{Mathematical description}

\textit{Notation.} We denote phase-space coordinates as a matrix $x \in \mathbb{R}^{6 \times N}$. Here the $n$-th column of $x$ contains all coordinates of particle $n$, i.e., $x_{:,n} = ({r}_n^x, {r}_n^y, { r}_n^z, {p}_n^x, {p}_n^y, {p}_n^z)^\mathrm{T}$, where ${ r}_n^k$ and ${ p}_n^k$ are components of the vectors ${\bf r}_n$ and ${ \bf p}_n$, respectively. Similarly, the diameters of the particles are given by $\sigma = (\sigma_1,\dots,\sigma_N) \in  \mathbb{R}^{1 \times N}$. In the following, we consider an arbitrary initial configuration $x_0$.

\textit{Transpositions $\tau_{ij}$.} Starting from $x_0$, each SMC algorithm below subsequently performs transpositions. A transposition $\tau_{ij}$ of particles $i \neq j$ is defined as
\begin{align}
     \big( \tau_{ij}(x) \big)_{:,n}
    = \begin{cases}
        x_{:,i} & \mathrm{if}~n=j,\\
        x_{:,j} & \mathrm{if}~n=i,\\
        x_{:,n} & \mathrm{otherwise}.\\
    \end{cases}
    \label{eq:transposition_definition}
\end{align}
This corresponds to an exchange of columns $i$ and $j$ in a configuration $x$. Trivially, a transposition conserves the total momentum. 

\textit{Permutations $\pi$.} From an algebraic perspective, a composition of transpositions is a permutation $\pi$. The reverse is also true: each permutation can be written as the composition of transpositions, such that the set $\mathcal{P}$ of all permutations is given by
\begin{align}
    \mathcal{P} = \left\{  \prod_{k=1}^K \tau_{i_k j_k} ~ | ~K \in \mathbb{N}, \,i_k \neq j_k \in \left\{ 1,\dots,N \right\} \right\},
    \label{eq:Permutations_as_Transpositions}
\end{align}
with
\begin{equation}
   \prod_{k=1}^K \tau_{i_k j_k} = \tau_{i_K j_K}  \circ \dots \circ \tau_{i_1 j_1}.
\end{equation}
$\mathcal{P}$ defines a group where the group operation is the composition, $\circ$.
The number of elements in $\mathcal{P}$ is $|\mathcal{P}| = N!$.

\textit{Symmetry.} A permutation of the coordinates $x$ is equivalent to the inverse of the same permutation of the diameters $\sigma$. This symmetry can be formulated in terms of the Hamilton function $H$ as
\begin{align}
   H \big(\,\pi(x)\,|\,\sigma\,\big) 
   = 
   H\big(\,x\,|\,\pi^{-1}(\sigma)\,\big),&&\forall \pi \in \mathcal{P}.
   \label{eq:symmetry_hamiltonian_SMC}
\end{align}
This identity can be verified as follows: Applying a permutation $\pi$ simultaneously to the coordinates $x$ as well as the diameters $\sigma$ is just a re-labeling of the particles. Thus, we have $H\left(\pi(x)|\pi(\sigma)\right) =  H\left(x|\sigma\right)$, from which Eq.~(\ref{eq:symmetry_hamiltonian_SMC}) follows. Note that here we assume identical particle masses in the model definition. Otherwise one had to incorporate the different masses into a generalized parameter matrix $\sigma \in \mathbb{R}^{2 \times N}$. Equation (\ref{eq:symmetry_hamiltonian_SMC}) implies that in simulations computationally efficient diameter exchanges can be used (right-hand side of Eq.~(\ref{eq:symmetry_hamiltonian_SMC})), but we can interpret their effect in terms of a sampling from a phase space at a fixed $\sigma$ (left-hand side of Eq.~(\ref{eq:symmetry_hamiltonian_SMC})).

\textit{Phase space $\Gamma_{x_0}$.} By sequentially applying swap moves, we sample from a discrete phase space $\Gamma_{x_0}$ and \textit{eventually} visit the set of all $N!$ possible permutations $\pi$ of the configuration $x_0$:
\begin{align}
    \Gamma_{x_0} &= \left\{ \pi(x_0),~\pi \in \mathcal{P}  \right\}. \label{eq:phase_space_discrete}
\end{align}
Note that $\Gamma_{x_0}$ has exactly the same number of elements as $\mathcal{P}$, since ${\bf r}_i \neq {\bf r}_j$ for $i \neq j$ is guaranteed by the pair interaction potential $u$. We emphasize that $\Gamma_{x_0}$ is only a subset of the total phase space of the hybrid MD-SMC dynamics.

\textit{Target distribution $W$.} The Metropolis-Hastings algorithm involves a target distribution $W$. We impose the canonical distribution with a temperature parameter $T_\mathrm{SMC}$ on the phase space $\Gamma_{x_0}$, i.e.,
\begin{align}
    W(x) &= Z^{-1}_{x_0} e^{- H(x|\sigma)/(k_B T_\mathrm{SMC})}, && x \in \Gamma_{x_0},
    \label{eq:phase_space_density_SMC}\\
    Z_{x_0} &= 
    \sum_{\pi \in \mathcal{P}} e^{-H(\pi(x_0)|\sigma)/(k_B T_\mathrm{SMC})}
    \label{eq:partition_sum_SMC}.
\end{align}
Here $Z_{x_0}$ is the partition sum with respect to the frozen configuration $x_0$. Note that for the sake of readability we leave out the index $x_0$ from the target distribution $W$ and other expressions below.

\textit{Metropolis-Hastings algorithm.} SMC uses the Metropolis-Hastings algorithm \cite{robert1999monte} to construct a Markov chain $(x_0, x_1, x_2, \dots)$. In each of the following SMC schemes, the same target distribution $W$ (defined above) is used, starting from the configuration $x_0$. Assume a configuration $x_{n-1}$ at ``time'' step $n-1$. To obtain the next configuration $x_{n}$, one first generates a trial configuration $x_*$ from a \textit{proposal probability} $q(\,.\,|x_{n-1})$. The choice of $q(\,.\,|\,.\,)$ defines the different SMC variants, discussed below. The trial configuration $x_*$ is accepted with a probability $\alpha(x_{n-1},x_*)$. If it is accepted, $x_n := x_*$, otherwise it is rejected, setting $x_n := x_{n-1}$. Here, the \textit{acceptance probability} is defined as
\begin{align}
    \alpha(x,y) = \mathrm{min}\left( 1, \frac{W(y)q(x|y)}{W(x)q(y|x)} \right).  
      \label{eq:metropolis_hastings_P_accept}
\end{align}
We will show that the proposal probability $q$ for each of the SMC schemes considered below is symmetric, i.e., $q(x |y ) =  q( y | x )$, so that the acceptance probability $\alpha$ simplifies to the special case of the \textit{Metropolis criterion},
\begin{align}
\alpha(x,y) = \mathrm{min}\left( 1, e^{-[H(y|\sigma)-H(x|\sigma)]/(k_B T_\mathrm{SMC})} \right).  
      \label{eq:metropolis_hastings_P_accept_simplified}
\end{align}

\textit{Convergence.} Now we discuss the conditions under which the SMC algorithm converges to the target distribution $W$. The \textit{transition probability} $P(y|x)$ to migrate from a state $x$ to a state $y$ is a conditional probability,
\begin{align}
    P(y|x) = \alpha(x,y) q(y|x).
    \label{eq:transition_probability}
\end{align}
Since the state space $\Gamma_{x_0}$ is finite, $|\Gamma_{x_0}| = N! < \infty$, $P(y|x)$ can be identified with a finite-dimensional transition matrix $\mathbf{P}$. The matrix notation demands that we count the state space and uniquely identify each state with one of these numbers. Similarly, let us denote the $n$-step transition probability to migrate from state $x$ to $y$ after $n$ steps by $P^{(n)}(y|x)$. Here the \textit{Kolmogorov-Chapman equation}~\cite{gusak2010theory} hold for all $n,\,m \geq 0$,
\begin{equation}
    P^{(n+m)}(y|x) = \sum_{z \in \Gamma_{x_0}} P^{(m)}(y|z) P^{(n)}(z|x).
    \label{eq:Kolmogorov_Chapman_Eqs}
\end{equation}
They can be used to express $P^{(n)}(y|x)$ by the associated entry of the matrix $\mathbf{P}^n$.
By construction, the Metropolis-Hastings method satisfies the \textit{detailed-balance condition}
\begin{align}
    W(x) P(y|x) = W(y) P(x|y),
\end{align}
which is also called \textit{reversibility of the chain}. It guarantees that $W$ is a \textit{stationary distribution} of the Markov chain, in the sense that
\begin{align}
    \sum_{x \in \Gamma_{x_0}}  W(x) P(y|x) = W(y).
\end{align}
In general, the existence of a stationary distribution itself is not sufficient to imply convergence. Before we state a theorem of convergence, we first need to define the properties \textit{aperiodicity} and \textit{irreducibility}.

\textit{Aperiodicity.} If a return to state $x$ can only occur in a multiple of $k$ steps, then $x$ is said to have a period of $k$. The period $k$ of a state $x$ is formally defined as
\begin{align}
    k(x) = \mathrm{gcd} \left\{ n \geq 1~|~ P^{(n)}(x|x) > 0 \right\},
    \label{eq:aperiodicity_Markov}
\end{align}
where $\mathrm{gcd}$ denotes the greatest common divisor. The Markov chain is said to be aperiodic if at least one state $x$ is aperiodic in the sense that $k(x) = 1$.

The Metropolis-Hastings algorithm with (i) a canonical distribution $W$ [cf.~Eqs.~(\ref{eq:phase_space_density_SMC}) and (\ref{eq:partition_sum_SMC})] as a target distribution and with (ii) a symmetric proposal probability $q$ is also called Metropolis algorithm. In this case, the Markov chain is aperiodic under very weak physical conditions: Consider that \textit{any} state $x \in \Gamma_{x_0}$ exists from which a state $x_*(x)$ with a higher energy can be proposed. Formally, this means that we assume $q(x_*(x)|x) > 0$ and $H(x|\sigma) < H(x_*(x)|\sigma)$. Then, there is a finite probability to reject $x_*$, cf.~Eq.~(\ref{eq:metropolis_hastings_P_accept_simplified}), since $\alpha(x, x_*) < 1$. Therefore, $P^{(1)}(x|x) \geq (1-\alpha(x,x_*))q(x_*|x) > 0$ and thus $k(x) = 1$.

\textit{Irreducibility.} A state $y$ is said to be \textit{accessible} from another state $x$ if the probability to transition from $x$ to $y$ in a finite number of steps is finite, i.e., if an integer $k(x,y) \geq 0$ exists with $P^{(k)}(y|x) > 0$. If $x$ is accessible from $y$ and $y$ from $x$, then both states are said to \textit{communicate}.
Communication defines an equivalence relation, whereby the maximal sets of communicating states represent equivalence classes.
A Markov chain is said to be \textit{irreducible} if each state communicates with each of the other states, i.e., the whole state space is one communicating class.

\textit{Ergodic theorem}~\cite{gusak2010theory}. Assume that a Markov chain is irreducible, aperiodic, and that a stationary distribution $W$ exists. Then, $W$ is the only stationary distribution and the Markov chain converges to $W$, in the sense that
\begin{align}
    \lim_{n \to \infty} P^{(n)}(y|x) = W(y).
    \label{eq:weak_convergence_markov_chain}
\end{align}
This type of \textit{weak convergence} states that, no matter in which state $x$ we currently are, after a sufficient number of steps we reach any state $y$ with probability $W(y)$.

In the following, three SMC schemes are discussed. For each variant, we introduce the algorithm, calculate its corresponding proposal probability $q$, and show that $q$ is symmetric. The question of convergence comes down to whether the Markov chain is irreducible, since the Metropolis method ensures that (i) the target distribution $W$ is a stationary distribution, and (ii) aperiodicity is guaranteed except for trivial configurations $x_0$.

\begin{figure*}
\includegraphics[scale=0.68]{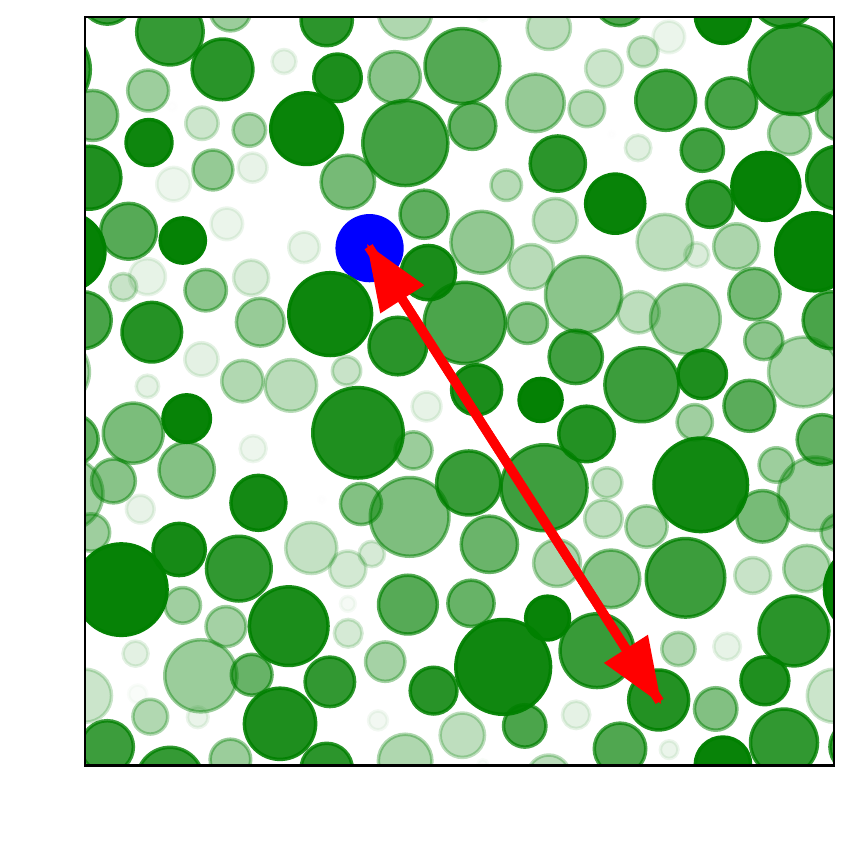}
\includegraphics[scale=0.68]{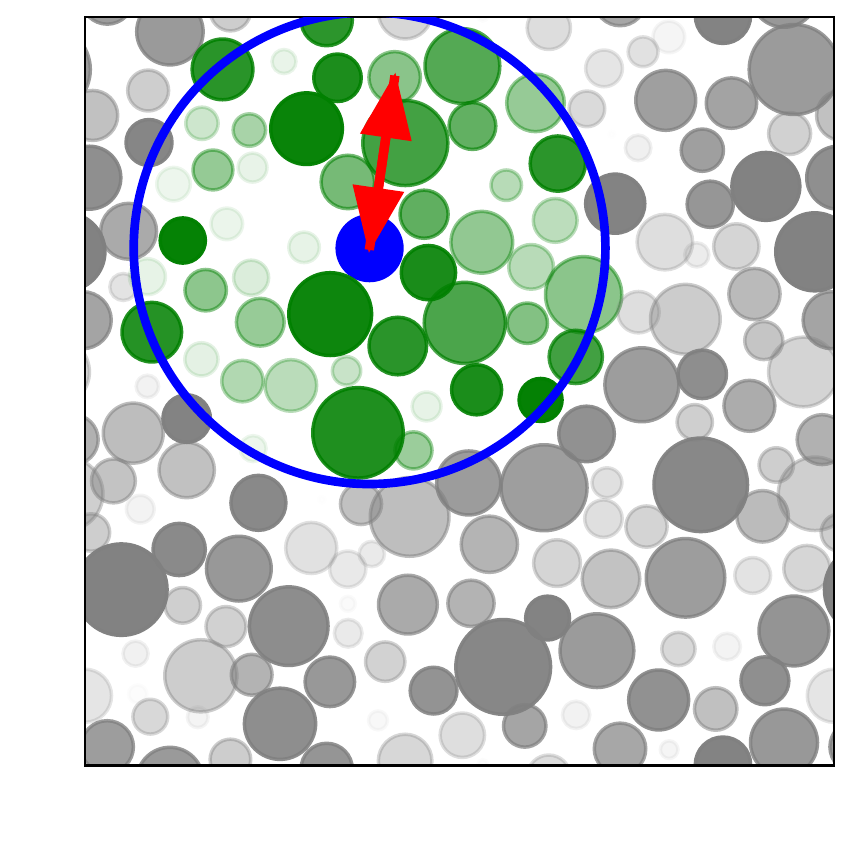}
\includegraphics[scale=0.68]{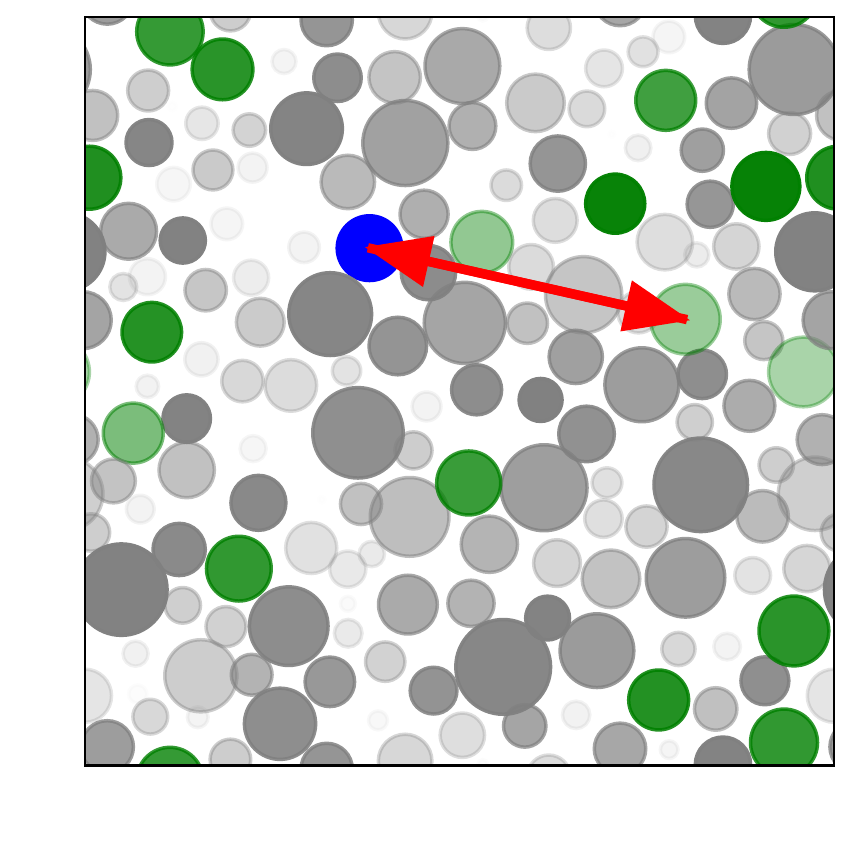}
\caption{Illustration of three SMC algorithms: standard (left), local (mid), and size-bias SMC (right panel). For a given blue particle, the green particles represent all of its \textit{allowed} transposition candidates. One of the possible swap attempts is indicated by a red arrow. The grey particles are not allowed to be exchanged with the blue one. For the standard SMC, transpositions between all particle pairs are considered. For the local SMC, only neighbors of the blue particle are allowed candidates, as indicated by the open blue circle. For the size-bias SMC, only particles with a similar diameter are considered. \label{Fig:SMC_variants_illustration}}
\end{figure*}
\subsection{Standard SMC\label{sec:ordinary_SMC}}
The standard SMC, to be introduced below, was first applied to binary mixtures \cite{tsai1978structure,gazzillo1989equation,kranendonk1991computer,grigera2001fast}. In the case of a polydisperse system, one randomly chooses a particle pair $(i,\,j)$ from a given configuration $x \in \Gamma_{x_0}$ and attempts a transposition to obtain the trial configuration $x_* = \tau_{ij}(x)$. Since each transposition is attempted with the same probability, the proposal probability $q( . | x )$ is a uniform distribution on the space of transpositions of $x$,
\begin{equation}
    q( y | x) = 
    \begin{cases}
        \frac{2}{N(N-1)}, &\mathrm{if}~i \neq j~\mathrm{exist}: y\equiv \tau_{ij}(x),\\
        0, &\mathrm{otherwise}.
    \end{cases}
    \label{eq:proposal_probability_ordinary_SMC}
\end{equation}
The symmetry $q(y|x) = q(x|y)$ holds since $\tau_{ij}(\tau_{ij}(x)) = x$.

In the following we show that the Markov chain of the standard SMC converges to the target distribution $W$. First, we note that the Markov chain is \textit{aperiodic} (except for trivial energetically degenerated configurations $x_0$, as discussed before) and that $W$ is a \textit{stationary distribution}. Both statements hold since SMC is a Metropolis algorithm. As a next step, irreducibility will be shown; then all three conditions of the ergodic theorem above are satisfied, so that convergence toward $W$ is established in the sense of Eq.~(\ref{eq:weak_convergence_markov_chain}).

\textit{Proof of irreducibility.} Let $x,\,y \in \Gamma_{x_0}$ be any configurations from the phase space of particle permutations, as introduced above. Since $y = \pi_1(x_0)$ and $x = \pi_2(x_0)$ for some $\pi_1,\,\pi_2 \in \mathcal{P}$, we can write $y = \pi(x)$ with $\pi = \pi_1 \circ \pi_2^{-1}$. According to group theory, every permutation can be written as a composition of transpositions. This means $\pi = \prod_{k=1}^K \tau_{i_k j_k}$ for a $K \in \mathbb{N}$ and transpositions $\tau_{i_k j_k}$. Let us now recursively define $z_k := \tau_{i_k j_k}(z_{k-1})$ with $z_0 := x$. This implies $z_K \equiv y$. From the Kolmogorov-Chapman Eq.~(\ref{eq:Kolmogorov_Chapman_Eqs}), the inequality $P^{(K)}(y|x) \geq \prod_{k=1}^K P( z_k | z_{k-1} )$ follows, which describes that the probability to transition from $x$ to $y$ in $K$ steps along \textit{any} path is greater than (or equal to) the probability to transition along \textit{the unique} path specified by the $K$ transpositions. Now, by definition~(\ref{eq:transition_probability}), we express the one-step transition probability $P( z_k | z_{k-1} ) = \alpha( z_{k-1} | z_k ) q( z_k | z_{k-1} )$ via the acceptance probability (\ref{eq:metropolis_hastings_P_accept_simplified}) and proposal distribution (\ref{eq:proposal_probability_ordinary_SMC}). Both probabilities are finite; it is $q( z_k | z_{k-1} ) > 0$ because $z_k = \tau_{i_k j_k}(z_{k-1})$ is a transposition of $z_{k-1}$. Therefore, we have $P^{(K)}(y|x) > 0$, i.e., the state $y$ is accessible from $x$. Since the states $x,\,y$ are arbitrary, the Markov chain is irreducible by definition.

%
\subsection{Local SMC\label{sec:local_SMC}}
The local SMC was introduced by Fernandez {\it et al.}~\cite{fernandez2007optimized}. The idea of this method is to only exchange particles for which the distance $r_{ij} \equiv |{\bf r}_i - {\bf r}_j|$ is smaller than a parameter $\Delta r > 0$. Given a configuration $x \in \Gamma_{x_0}$, let $\mathcal{N}(x)$ denote the list of all particles which have at least one neighbor,
\begin{align}
    \mathcal{N}(x) = \left\{~i = 1,\dots,N~ |~ \exists j \neq i : r_{ij} < \Delta r ~\right\}.
\end{align}
For $\Delta r \gtrsim 1$ and dense liquid samples with number density $N/V = 1$ considered in our work, we have $\mathcal{N}(x) = \{1,\dots,N\}$ for all typical configurations. Analogously, let $\mathcal{N}_i(x)$ be the list of all the neighbors of a particle $i$,
\begin{align}
    \mathcal{N}_i(x) = \left\{~j = 1,\dots,N~ |~j\neq i,~r_{ij} < \Delta r ~\right\}.
\end{align}
In the trivial case where no neighboring particles at all exist, $|\mathcal{N}(x)| = 0$, we propose $x_* = x$ such that $q(x|x) = 1$. In the non-trivial case, the local SMC algorithm first randomly picks a particle $i \in \mathcal{N}(x)$, subsequently chooses a random neighbor $j \in \mathcal{N}_i(x)$, and then proposes the transposition $x_* = \tau_{ij}(x)$. The corresponding proposal probability is
\begin{align}
    q(y|x) =
    \begin{cases}
        \frac{1}{|\mathcal{N}(x)|} \left(\frac{1}{ |\mathcal{N}_i(x)|} + \frac{1}{|\mathcal{N}_j(x)|}\right),
      &\mathrm{if}~\exists\,i \in \mathcal{N}(x)\\
      &\land~\exists j \in \mathcal{N}_i(x):\\
      &y \equiv \tau_{ij}(x),\\
         0, &\mathrm{otherwise}.
    \end{cases} 
    \label{eq:definition_proposal_probability_local_SMC}
\end{align}
The sum in the first row accounts for the two possibilities by which a transposition $\tau_{ij}(x)$ can be proposed with the algorithm: One option is to first choose the particle $i$ and then to pick $j \in \mathcal{N}_i(x)$. The second option is to first choose $j$ and then select $i \in \mathcal{N}_j(x)$.

One can show the symmetry $q(y|x) = q(x|y)$ as follows: The neighbors of particle $i$ in configuration $x$ are the same as the neighbors of particle $j$ in configuration $y = \tau_{ij}(x)$, since $j$ now occupies the former coordinates of $i$. This means that $\mathcal{N}_j(y) = \mathcal{N}_i(x)$. Similarly, the particles with neighbors remain the same, i.e., $\mathcal{N}(x) = \mathcal{N}(y)$.

Regarding the question of convergence, in the most general case, irreducibility of the SMC Markov chain depends on the relation between the swap range $\Delta r$ and the frozen configuration $x_0$ (and thus the density and the temperature at which $x_0$ was prepared). In the case that $\Delta r$ is too small (smaller than the typical distance between neighboring particles), there might exist distinct communicating classes (separated clusters of particles), between which particles cannot be exchanged. 

A sufficient condition for the convergence of the local SMC, which should be satisfied in typical configurations of dense liquids, can be obtained as follows. We assume that for any two particles $(i,\,j)$ we can find a path from $i$ to $j$ along a chain of neighboring particles. Formally, this means that particle indices $i_k$ exist such that ${\bf r}_i - {\bf r}_j = \sum_{k=1}^K {\bf r}_{i_k} - {\bf r}_{i_{k+1}}$ is a telescoping sum with $|{\bf r}_{i_k} - {\bf r}_{i_{k+1}}| < \Delta r$ for all $k = 1,\,\dots,\,K$. Note that $i_1 \equiv i$ and $i_{K+1} \equiv j$. Then, the transposition $\tau_{ij}$ can be obtained by a sequence of swaps between neighboring particles, $\tau_{ij} = \left(\prod_{k=1}^{K-1} \tau_{i_{K-k} i_{K-k+1}}\right) \circ \left(\prod_{k=1}^{K} \tau_{{i_k} i_{k+1}}\right)$.
This expression means that we first sequentially swap particle $i$ until we reach $j$ and then reversely swap $j$ along the same path to the former position of $i$.
Now that we can realize \textit{any} transposition with a finite sequence of the local SMC, also \textit{any} permutation can be obtained, cf.~Eq.~(\ref{eq:Permutations_as_Transpositions}). Thus we can transition from any state to any other in a finite number of steps with a finite probability (for more mathematical details, see the proof of irreducibility for the standard SMC). Hence the Markov chain is irreducible by definition and it converges according to the ergodic theorem above in the sense of Eq.~(\ref{eq:weak_convergence_markov_chain}).

From the local SMC, the standard SMC can be recovered in the limit $\Delta r \to \infty$. In our simulations the algorithms are identical if $\Delta r \geq L \sqrt{3}/2$.

In the introduction of the local SMC algorithm in Ref.~\cite{fernandez2007optimized}, an erroneous assumption about the proposal probability $q$ was made. Here the authors did not take into account the second summand in the first row of Eq.~(\ref{eq:definition_proposal_probability_local_SMC}), and thus the proposal probability $q$ that they assumed was not symmetric. With this $q$, an incorrect expression for the acceptance probability was obtained. 


The local SMC might be well suited as a potential candidate for a parallelized implementation of SMC.


%
\subsection{Size-bias SMC\label{sec:size-bias_SMC}}
The size-bias SMC was introduced in a work by Brumer and Reichman \cite{brumer2004numerical} who referred to this method as ``swap-sector Monte Carlo''. The idea of this variant is to avoid attempts of transpositions which are rejected with a high probability due to a large difference between the diameters. The anatomy of the size-bias SMC is similar to the local SMC, except that the metric to identify ``neighboring'' particles is applied within the diameter space with a cutoff $\Delta \sigma > 0$. Formally, we adopt exactly the same algorithm and definitions as in Sec.~\ref{sec:local_SMC}, but we replace $r_{ij}$ by $|\sigma_i - \sigma_j|$ and $\Delta r$ by $\Delta \sigma$. Note that we do not take into account the non-additivity of the diameters in the calculation of $\Delta \sigma$. 

While the convergence of the local SMC generally depends on the configuration $x_0$, the convergence of the size-bias Markov chain toward the target distribution $W$ can always be ensured by the choice of a sufficiently large system size $N$: again, the question of convergence boils down to whether all states communicate with each other (irreducibility). To this end, we have to show that we can realize any permutation with the size-bias SMC. Here, the argumentation is as follows. As before, we only need to show that \textit{any} transposition $\tau_{ij}$ between any two particles $(i,\,j)$ can be realized, because each permutation can be written as a composition of transpositions, see Eq.~(\ref{eq:Permutations_as_Transpositions}). The apparent problem for the size-bias SMC is, however, that only transpositions between \textit{similar} diameters are allowed. To this end, let us first assume, without loss of generality, that the diameters $\sigma_k$ are sorted in any order. For any given $\Delta \sigma >0$, we can choose a sufficiently large system size $N$ such that $|\sigma_{k+1} - \sigma_k| < \Delta \sigma$ for all $k=1,\dots,N-1$. This means that each particle has a smaller and a larger ''neighbor'' within the cutoff $\Delta \sigma$, except for the boundary particles. Note that this assumes a deterministic method~\cite{kuchler2022choice} to choose the diameters in the polydisperse model, as well as a compact domain of the diameter distribution density $f$, cf.~Sec.~\ref{sec:interaction_model}. Then, each transposition $\tau_{ij}$ can be obtained by swapping sequentially \textit{only} between particles with a similar diameter: $\tau_{ij} = \left(\prod_{k=1}^{j-i-1} \tau_{(j-k)j}\right) \circ \left(\prod_{k=i+1}^{j} \tau_{ik}\right)$.

An efficient implementation of the proposal part of the size-bias SMC with pseudo code reads:

%
\definecolor{backcolor}{rgb}{0.95,0.95,0.82}
\begin{lstlisting}[language=C++,tabsize=1,basicstyle=\small,
keywordstyle=\color{Brown}\textbf,commentstyle=\color{blue},
numbers=right, numbersep=1pt, columns=fullflexible,
emph={i,j},emphstyle=\color{Magenta}\textbf,
backgroundcolor=\color{backcolor}]
i = RNDM_INTEGER( generator ); // random particle p[i]
while( true ){
    j   = RNDM_INTEGER( generator );
    dij = fabs( p[i].sigma - p[j].sigma ) // distance i to j
    if( dij < DS  &&  j != i ) break;
}
\end{lstlisting}
This short code snippet illustrates the simplicity of the algorithm: in comparison to the standard SMC, we only add the calculation of $|\sigma_i - \sigma_j|$ alias ``dij'' in line 4 and a float comparison ``dij $<$ DS'' in line 5.
   
For our model, the size-bias SMC outperforms the local as well as the standard SMC, as we will show in Sec.~\ref{Sec:numerical_results_swap_on_a_grid}. 


\section{Numerical results for SMC\\on a frozen configuration \label{Sec:numerical_results_swap_on_a_grid}}

In this section we consider swap Monte Carlo (SMC) on a frozen configuration $x_0$ and evaluate the performance of the three SMC schemes introduced above. To this end, we determine the acceptance rate $P_\mathrm{accept}$, a diameter correlation function $C_\sigma$, and a relaxation time $s^\mathrm{rel}$. For an initial configuration $x_0$ that was equilibrated at a specific temperature $T$ before, cf.~Sec.~\ref{sec:equilibration_protocol}, we now apply SMC at the same temperature, i.e., $T_\mathrm{SMC} = T$ in Eq.~(\ref{eq:metropolis_hastings_P_accept_simplified}). In this sense, the numerical results of this section can be transferred to the \textit{equilibrium} simulations with hybrid MD-SMC in Sec.~\ref{sec:MD_SMC}.

\begin{figure}
\includegraphics{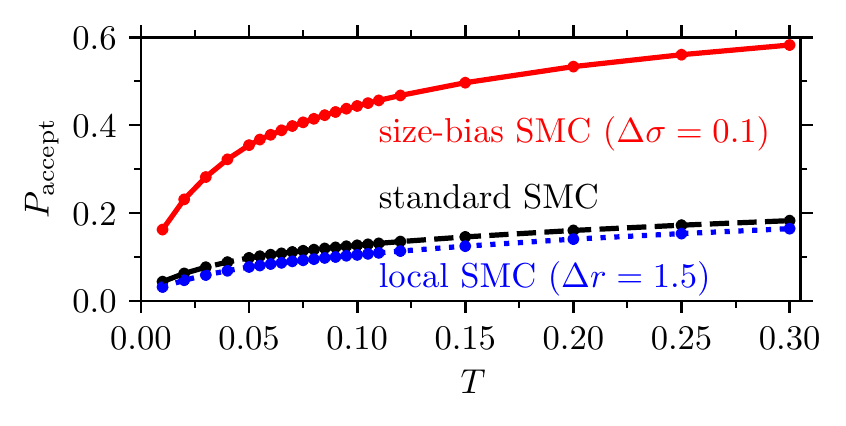}
\caption{Probability $P_\mathrm{accept}$ of accepting a trial diameter exchange as a function of temperature $T$ for three SMC schemes. \label{fig_acceptance_rate_swap}}
\end{figure}
\textit{Acceptance rate $P_\mathrm{accept}$.} In Fig.~\ref{fig_acceptance_rate_swap}, we show the acceptance rate $P_\mathrm{accept}$ of trial swaps as a function of temperature $T$. Here, $P_\mathrm{accept}$ is calculated by dividing the number of accepted attempts $x_*$ by the total number of attempts of a Markov chain of length $10^3 \times N$, averaged over $60$ initial configurations $x_0$ with $N$ particles each. Under equilibrium conditions, i.e., for temperatures $T > T_\mathrm{g}^\mathrm{SMC} \approx 0.06$, the acceptance rates are above 8.4\% for all three specified SMC methods. 

In the work of Fernandez {\it et al.}~\cite{fernandez2007optimized}, local SMC for a binary system is proposed, resulting in a larger acceptance rate $P_\mathrm{accept}$ than for the standard SMC. In contrast, for our polydisperse model, the local SMC with $\Delta r = 1.5$ has a slightly smaller $P_\mathrm{accept}$ than the standard scheme. This qualitatively different behavior presumably results from a different chemical ordering in a binary and a polydisperse system. 

We want to emphasize that $P_\mathrm{accept}$ is not a suitable measure to compare the \textit{performance} of SMC algorithms. This will become clear for the size-bias SMC below: while decreasing $\Delta \sigma$ always leads to higher acceptance rates, the SMC moves between too similar diameters are inefficient. Instead, we will now propose a diameter relaxation function as a reasonable performance measure. From this correlation function, we shall infer that the local SMC is inferior to the standard SMC for any $\Delta r$ (for our model system at a low temperature $T = 0.065$).


%
\begin{figure}
\includegraphics{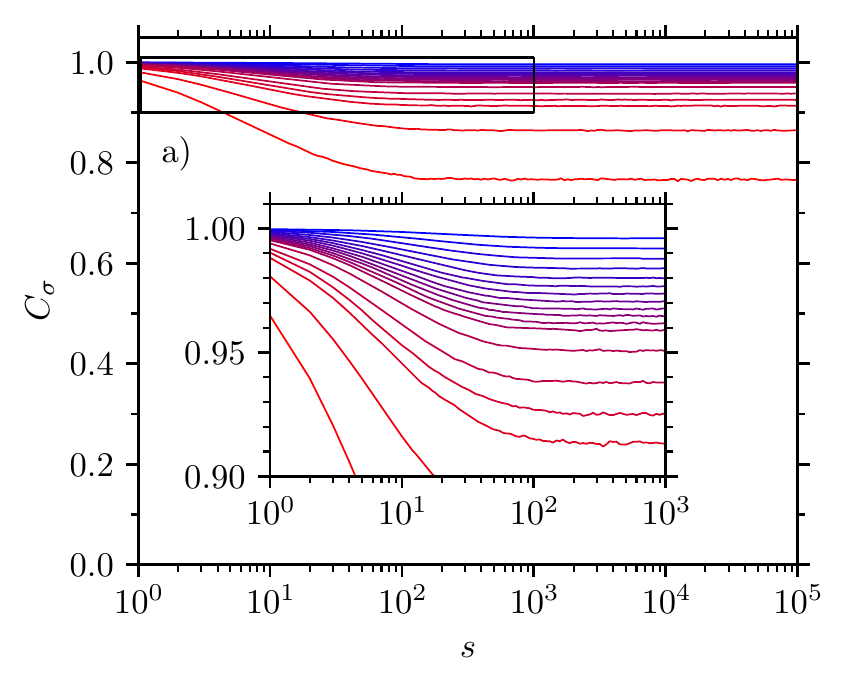}
\includegraphics{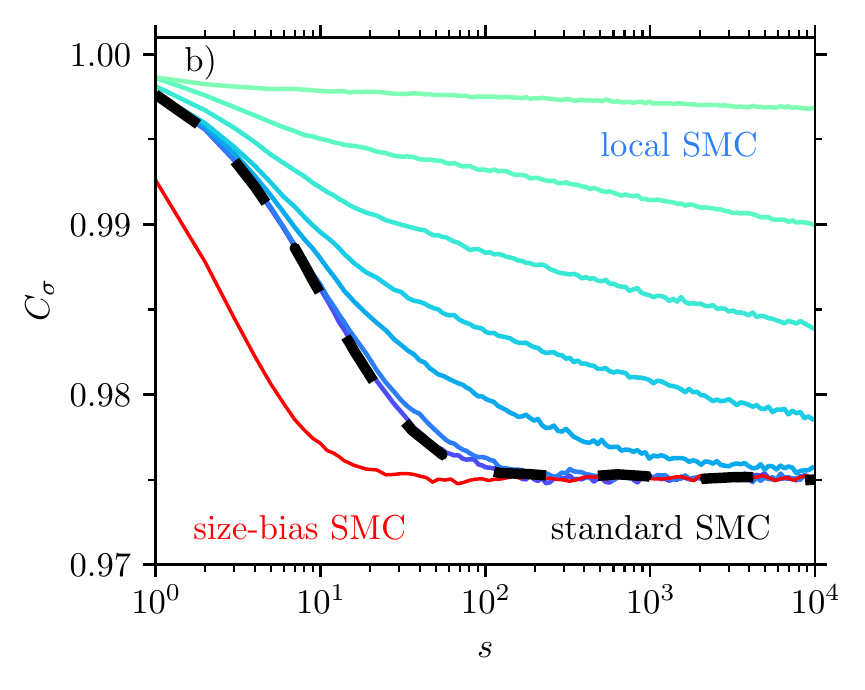}
\caption{Correlation function $C_\sigma$ as a function of time $s$ (corresponding to the number of swap sweeps) in frozen configurations, averaged over $60$ samples with $N=2048$ particles each. a) Standard SMC for different temperatures $T = 1$, $0.5$, $0.3$, $0.25$, $0.2$, $0.15$, $0.12$, $0.11$, $0.1$, $0.09$, $0.08$, $0.07$, $0.06$, $0.05$, $0.04$, $0.03$, $0.02$, $0.01$, changing the color with decreasing temperature from red to blue. b) At the fixed temperature $T=0.065$ for the standard SMC (black dashed line), the size-bias SMC with $\Delta \sigma = 0.1$ (red line), and the local SMC for $\Delta r = 1$, $1.25$, $1.5$, $1.75$, $2$, $3$, $4$ (with increasing $\Delta r$ the color of the curves changes from turquoise over blue to purple). \label{Fig:pure_SMC_C_D}}
\end{figure}

\textit{Diameter autocorrelation function $C_\sigma$.} An appropriate quantity to compare the performance of the different SMC methods is a diameter (auto-)correlation function $C_{\sigma}(s)$, applying swaps (permutations) of the diameters on fixed phase-space coordinates $x_0$, cf.~Eq.~(\ref{eq:symmetry_hamiltonian_SMC}). In order to measure the elapsed ``time'' $s$ in a system-size independent way, we use the number of swap sweeps; here one sweep is defined as $N$ elementary SMC trials. The function $C_{\sigma}(s)$ quantifies the time correlation of a diameter fluctuation $\sigma_i(s) - \sigma_\mathrm{av}$ around the average diameter $\sigma_{\rm av} = \frac{1}{N} \sum_{i=1}^N \sigma_i \approx \bar{\sigma}$. It is defined by 
\begin{equation}
    C_\sigma(s) = 
    \frac{\mathbb{E}\left[\sum_{i=1}^N (\sigma_i(s) - \sigma_\mathrm{av})(\sigma_i(0)-\sigma_\mathrm{av}) \right]}
    {\mathbb{E}\left[\sum_{i=1}^N (\sigma_i(0) - \sigma_\mathrm{av})^2  \right]}
   .
\end{equation}
Here, $\mathbb{E}[\,.\,]$ denotes an expectation value with respect to the Markov chain as well as the distribution of initial configuration $x_0$. In practice, we use only one realization of the Markov chain at a given $x_0$ and then average over the ensemble of $60$ independent samples $x_0$.

\begin{figure}
\centering
\includegraphics{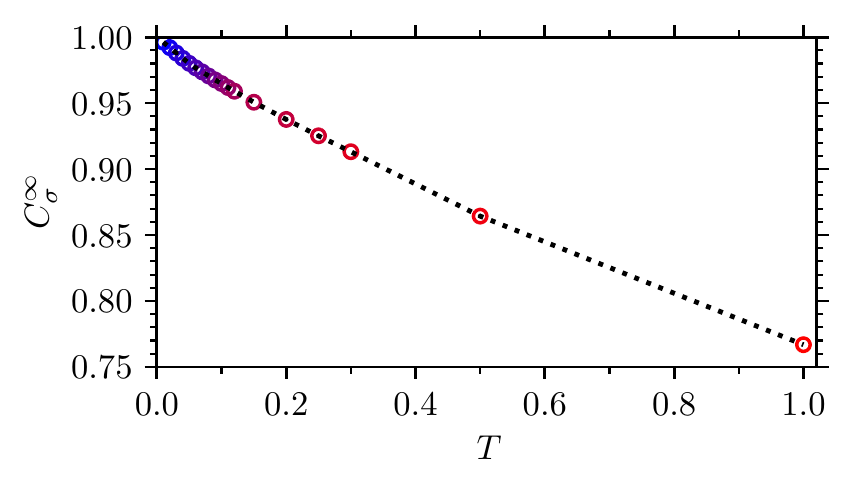}
\caption{Plateau $C_\sigma^\infty \equiv \lim_{s \to \infty} C_\sigma(s)$ as a function of $T$. \label{Fig:pure_SMC_C_plateau_T}}
\end{figure}
In Fig.~\ref{Fig:pure_SMC_C_D}a, we show $C_\sigma(s)$ for different temperatures $T$. One observes that in the long-time limit the correlation function decays onto a well-defined temperature-dependent plateau,
\begin{align}
    C_\sigma^\infty := \lim_{s \to \infty} C_\sigma(s) > 0.
\end{align}
A finite value means that the system keeps some memory of its initial diameter configuration forever. From a numerical perspective, large values $C_\sigma(s) > 0.75$, even at a very high temperature $T = 1$, imply that the ``allowed'' diameter fluctuations within a frozen configuration are rather small.

Figure~\ref{Fig:pure_SMC_C_D}b shows $C_\sigma(s)$ at a fixed temperature $T=0.065$ for the three SMC variants. For the size-bias SMC (red line, $\Delta \sigma = 0.1$), the function $C_\sigma(s)$ decays much faster than for the standard SMC (black dashed line), which in turn outperforms the whole set of local SMC algorithms ($\Delta r$ increases from turquoise over blue to purple lines). Upon increasing $\Delta r$ for the local SMC, $C_\sigma(s)$ continuously approaches that of the standard SMC. Already for $\Delta r = 3$ the curves of the local and standard SMC are very close. This finding is relevant with respect to a possible parallelization of SMC using the local SMC. Independent of the SMC variant, the function $C_\sigma(s)$ seems to approach the same plateau value $C_\sigma^\infty$ as $s \to \infty$; this numerical result is an implication of our analytical result from Sec.~\ref{Sec:theory_swap_on_a_grid}, that each SMC scheme converges to the same target distribution $W$ (under the conditions elaborated in Sec.~\ref{Sec:theory_swap_on_a_grid}).

Figure~\ref{Fig:pure_SMC_C_plateau_T} shows the plateau height $C_\sigma^\infty$ as a function of temperature $T$. The function $C_\sigma^\infty(T)$ is monotonically decreasing, which means that the lower $T$ is, the smaller is the ``accessible'' diameter space of each particle. This finding represents an analogy to the cage effect, where an increasing localization of the particles upon decreasing $T$ can be observed (see below).

\begin{figure}
    \includegraphics{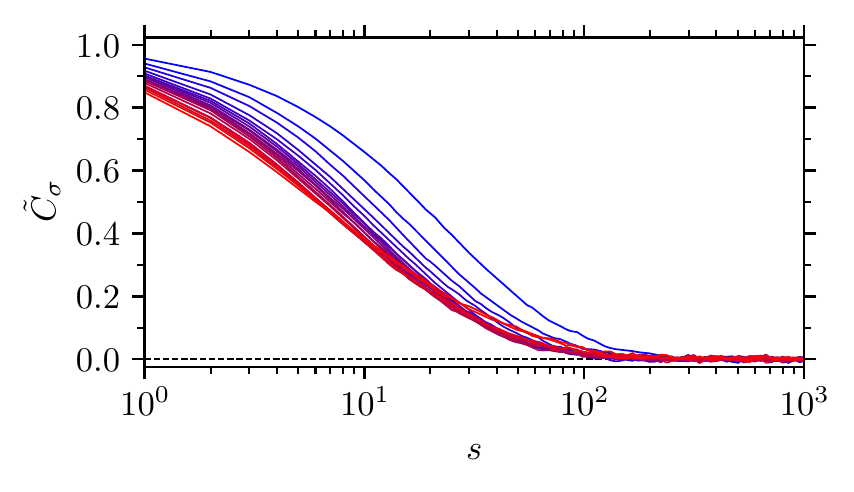}
	\caption{Correlation function $\tilde{C}_\sigma(s)$, see Eq.~(\ref{eq:C_sigma_t_plateau}), as a function of SMC sweeps $s$ in frozen coordinates. Temperatures $T$ increase from blue to red, see the caption of Fig.~\ref{Fig:pure_SMC_C_D}a. \label{Fig:pure_SMC_C_D_tilde}}
\end{figure}

\textit{Relaxation time $s^\mathrm{rel}$.} To quantify how fast SMC ``thermalizes'' the diameters, we first measure the decay of $C_\sigma(s)$ onto the plateau $C_\sigma^\infty$ with a normalized correlation function $\tilde{C}_\sigma(s)$, and then define a relaxation time $s^\mathrm{rel}$. To this end, we introduce
\begin{align}
	\tilde{C}_\sigma(s) = \frac{C_\sigma(s) - C_\sigma^\infty}{C_\sigma(0) - C_\sigma^\infty}.
 \label{eq:C_sigma_t_plateau}
\end{align}
In Fig.~\ref{Fig:pure_SMC_C_D_tilde}, we display $\tilde{C}_\sigma$ as a function of $s$ for the standard SMC at the same temperatures $T$ as in Fig.~\ref{Fig:pure_SMC_C_D}a.
We define a relaxation time $s^\mathrm{rel}$ via
\begin{align}
   \tilde{C}_\sigma(s^\mathrm{rel}) = e^{-1}.
\end{align}
If we ignore the lowest five temperatures $T < T_\mathrm{g}^\mathrm{SMC}$ corresponding to glassy non-equilibrium states, we find a relaxation time $s^\mathrm{rel} \approx 10$, almost independent of the temperature $T$.

\begin{figure}
\includegraphics{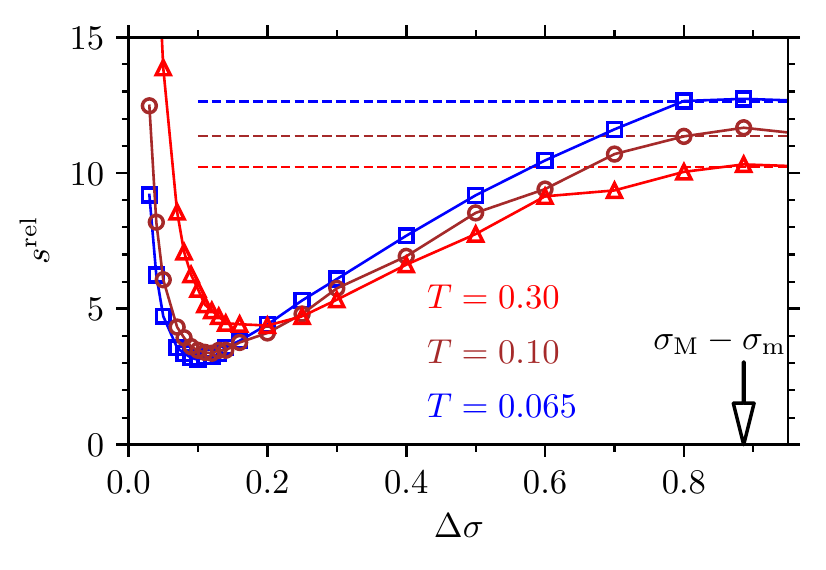}
\caption{Relaxation time $s^\mathrm{rel}$ as a function of the parameter $\Delta \sigma$ of the size-bias SMC for three different temperatures $T$. The dashed horizontal lines denote the relaxation times of the standard SMC. The black arrow marks $\sigma_\mathrm{M} - \sigma_\mathrm{m}$, the difference between the maximum and minimum diameter. \label{Fig:tau_ds_size-bias_SMC}}
\end{figure}

Figure~\ref{Fig:tau_ds_size-bias_SMC} shows the relaxation time $s^\mathrm{rel}$ as a function of the parameter $\Delta \sigma$ of the size-bias SMC for the temperatures $T=0.065$, 0.1, and 0.3. We observe that $s^\mathrm{rel}(\Delta \sigma)$ has a minimum, $\Delta \sigma _\mathrm{min}(T)$. For the specified temperature range we have $0.1 \lesssim \Delta \sigma _\mathrm{min} \lesssim 0.2$. Since we are interested in optimizing SMC at low temperatures close to the numerical glass-transition temperature $T_\mathrm{g}^\mathrm{SMC} \approx 0.06$, we propose $\Delta \sigma _\mathrm{min} = 0.1$ as the optimized value for this model system. The existence of a minimum for the size-bias SMC is intuitively clear: while too large $\Delta \sigma$ lead to ``unnecessary'' SMC trials which are rejected most of the times, too small values are also inefficient since then only very similar diameters are exchanged such that the swap moves have essentially no effect.

\textit{Computational efficiency.} Above, we have quantified the ``physical'' performance of the different SMC schemes in terms of the diameter correlation function $C_\sigma(s)$. However, this does not fully account for the computational efficiency, for which the computational load has to be considered as well. We find that the three considered SMC variants can be implemented efficiently, as we demonstrated with the code snippet in Sec.~\ref{sec:size-bias_SMC}. Their computational load is similar, as we will see below in Fig.~\ref{Fig:CPU_time} in Sec.~\ref{sec:MD_SMC}.

\section{Hybrid MD-SMC Dynamics\label{sec:MD_SMC}}

\textit{Definition and parameters.} In this section we analyze the hybrid MD-SMC dynamics, introduced in Ref.~\cite{berthier2019efficient}. This dynamics consists of microcanonical ($NVE$) molecular dynamics (MD) simulation where, periodically after a time interval $t_\mathrm{MD}$, $s$ consecutive swap sweeps are inserted. Here, one sweep is defined by $N$ subsequent elementary SMC attempts to exchange the diameters of particle pairs while the coordinates are frozen, as defined in Sec.~\ref{Sec:theory_swap_on_a_grid}. The numerical results presented below refer to the \textit{standard} SMC scheme, however the same results are expected for the other two variants. Their efficiency is different but with the choice of appropriate parameters they sample from the same target distribution.

Following Ref.~\cite{berthier2019efficient}, we define the SMC frequency
\begin{equation}
    f_\mathrm{SMC} = \frac{s}{t_\mathrm{MD}} \label{eq:f_SMC_density}
\end{equation}
to fully characterize MD-SMC by these three parameters, among which only two are independent. In the following, we will refer to MD-SMC with a specific choice of $t_\mathrm{MD}$ and $s$ as $\mathrm{SMC}(t_\mathrm{MD},s)$. By a comprehensive analysis of $\mathrm{SMC}(t_\mathrm{MD},s)$ with varying $t_\mathrm{MD}$ and $s$, we give insight into the mechanism of the drastically accelerated structural relaxation. Before that, we demonstrate that the system is properly thermostatted when coupled to SMC. Unless noted otherwise, we use $t_\mathrm{MD} = 0.01 \equiv \Delta t$ as a default value in the following.

\subsection{MD-SMC as a thermostat \label{sec:SMC_as_thermostat}}
The sole addition of SMC to microcanonical MD can be used to adjust the temperature $T$ of the system. Thus, it is not necessary to couple MD to a thermostat such as the Nos\'e-Hoover or the Berendsen thermostat \cite{allen2017computer}, or the Lowe-Andersen thermostat applied in our equilibration protocol, cf.~Sec.~\ref{sec:equilibration_protocol}. To see this, we perform the following protocol: We start from equilibrated configurations at the initial temperature $T_0 = 0.30$, followed by MD-SMC simulation at a target temperature $T_\mathrm{SMC}$ that enters the Metropolis criterion of the swap moves, cf.~Eq.~(\ref{eq:metropolis_hastings_P_accept_simplified}). We determine the instantaneous temperature $T := \langle \frac{2 K}{3 N}\rangle$, averaged over 60 simulations, via the kinetic energy $K$ of a sample with $N$ particles.

\begin{figure}
\includegraphics{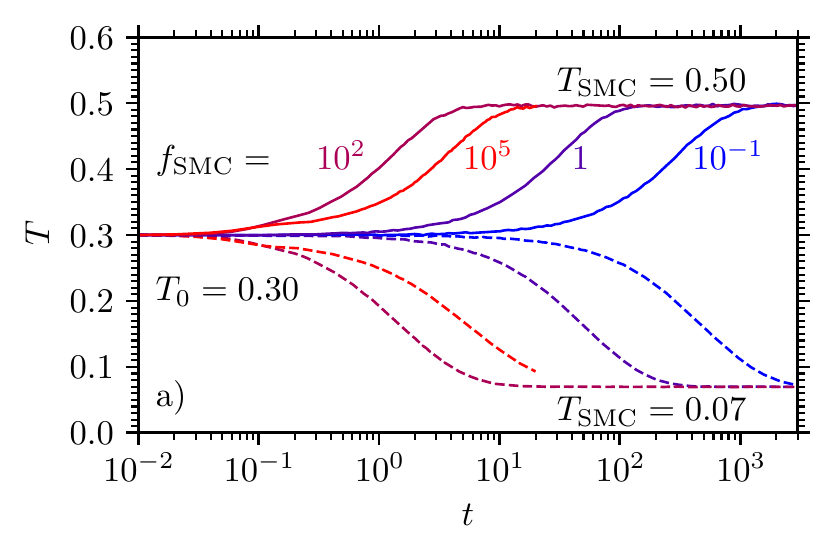}
\includegraphics{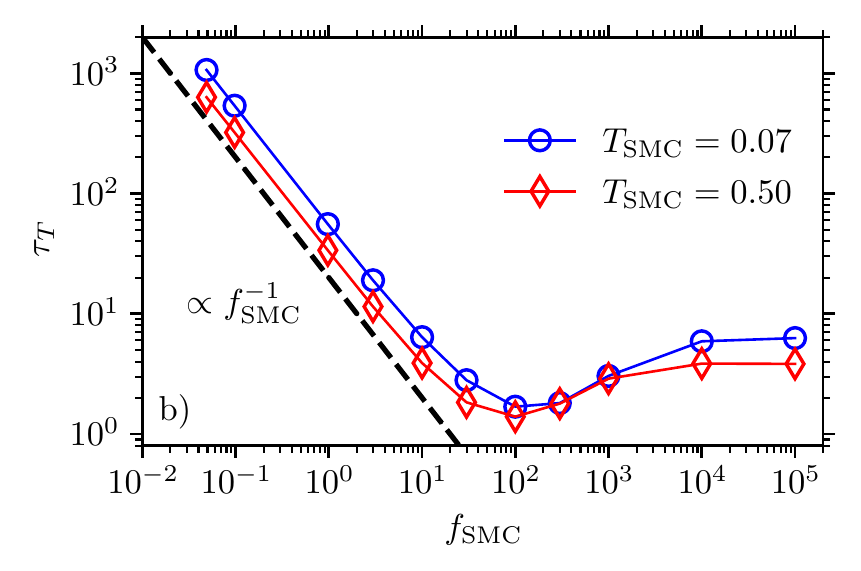}
 \caption{a) Instantaneous temperature $T$ as a function of time $t$ with MD-SMC dynamics ($t_\mathrm{MD} = 0.01$). The initial temperature is $T_0 = 0.30$ for each curve. Two different target temperatures $T_\mathrm{SMC}$ are imposed (cf.~Metropolis criterion of swap trial moves, Eq.~(\ref{eq:metropolis_hastings_P_accept_simplified})); results for $T_\mathrm{SMC} = 0.07$ are shown as dashed lines and $T_\mathrm{SMC} = 0.50$ as solid lines. The frequency $f_\mathrm{SMC}$ is varied for both $T_\mathrm{SMC}$, as indicated. b) Relaxation time $\tau_T$, see Eq.~(\ref{eq:SMC_cooling_rate}), as a function of frequency $f_\mathrm{SMC}$ for the two target temperatures $T_\mathrm{SMC}$ used in a). The dashed black line indicates a proportionality $\tau_T \propto f_\mathrm{SMC}^{-1}$. \label{Fig:SMC_as_thermostat}}
\end{figure}
Figure~\ref{Fig:SMC_as_thermostat}a shows the instantaneous temperature $T$ as a function of time $t$ for the target temperatures $T_\mathrm{SMC} = 0.07$ (dashed lines) and $T_\mathrm{SMC} = 0.50$ (solid lines), i.e., a very low and a relatively high temperature (see below). For the frequencies $f_\mathrm{SMC} \in \{ 10^{-1},\,1,\,10^2\}$, both target temperatures are approached on a time scale that decreases with increasing frequency $f_\mathrm{SMC}$. A non-monotonic behavior occurs for $f_\mathrm{SMC} = 10^5$, where the time scale increases again.

For a quantitative analysis, let us introduce the relaxation time $\tau_T$ of the temperature, defined by
\begin{equation}
    \frac{T(\tau_T) - T_\mathrm{SMC}}{T_0 - T_\mathrm{SMC}} =  {\rm e}^{-1}.
    \label{eq:SMC_cooling_rate}
\end{equation}
Figure~\ref{Fig:SMC_as_thermostat}b shows $\tau_T$ as a function of $f_\mathrm{SMC}$ for $T_\mathrm{SMC} = 0.07$ and $T_\mathrm{SMC} = 0.50$. The quantitative behavior is very similar in both cases. For $f_\mathrm{SMC}\lesssim 10$, we observe $\tau_T \propto f_\mathrm{SMC}^{-1}$. Then, upon increasing $f_\mathrm{SMC}$, the relaxation time $\tau_T$ reaches a shallow minimum at a value $\tau_T^{\rm min} \approx 2$ around $f_\mathrm{SMC}^{\rm min} \approx 10^2$. For larger $f_\mathrm{SMC}$, $\tau_T$ increases mildly and then saturates at the values $\tau_T \approx 6$ for $T_\mathrm{SMC} = 0.07$ and $\tau_T \approx 4$ for $T_\mathrm{SMC} = 0.50$. The saturation of $\tau_T$ can be understood from our findings for SMC on a frozen configuration in Sec.~\ref{Sec:numerical_results_swap_on_a_grid}. Here, we saw that about $s^\mathrm{rel} \approx 10$ swap sweeps are required for the decay of the diameter correlation function onto a plateau. By performing more than $s^\mathrm{rel}$ swap sweeps per microscopic time scale $t_\mathrm{mic} \approx 0.2$ (to be defined shortly), we expect a saturation of the MD-SMC dynamics with respect to the speed with which the target temperature is approached. The swap frequency at which this saturation threshold is approached can be estimated as $f_\mathrm{SMC}^* = s^\mathrm{rel}/t_\mathrm{mic} \approx 50$ which roughly corresponds to the ``onset'' of the shallow minimum in $\tau_T$. The reason for the increase from $f_\mathrm{SMC}^{\rm min}$ to the final saturation above $f_\mathrm{SMC}^{\rm min}$ is however not clear to us.

\begin{figure}
\includegraphics{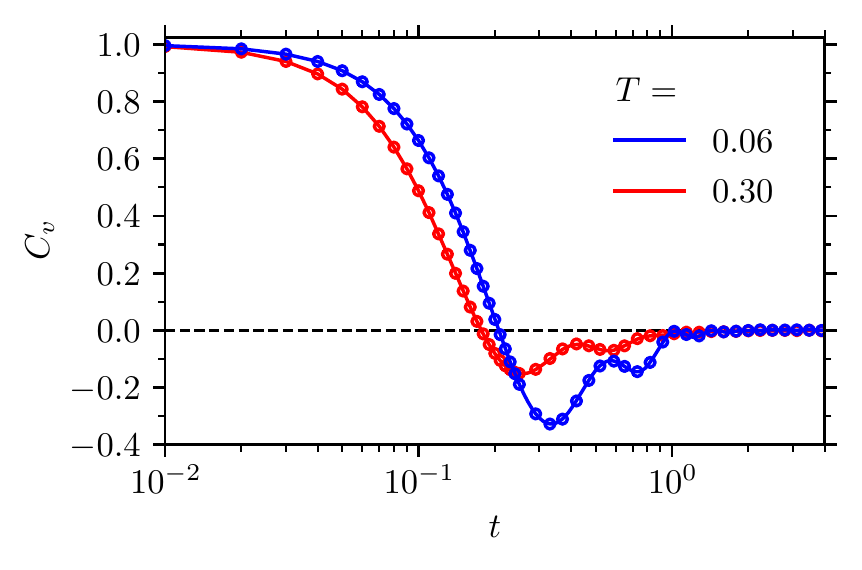}
\caption{Velocity correlation function $C_v(t)$ for $N=8000$ particles and the two temperatures $T=0.06$ and $T=0.30$, computed with pure $NVE$ dynamics (solid lines) and with $NVE$ dynamics \textit{subjected to a singular distortion} at $t=0$ (circles). The distortion is defined as a full thermalization of diameters by performing $10^3 \times N$ swap attempts.  \label{Fig:C_V_t}}
\end{figure}
A microscopic time scale $t_\mathrm{\rm mic}$ can be estimated via the first zero-crossing of the velocity autocorrelation function
\begin{equation}
    C_v(t) = \langle {\bf v}(t) {\bf v}(0) \rangle / \langle {\bf v}^2(0) \rangle.
\end{equation}
Thus, the microscopic time scale is given by $t_\mathrm{mic} = \min \{ t\,|\,C_v(t)=0\}$. As we shall see below, a similar estimate of $t_\mathrm{mic}$ is the location of the maximum of the derivative of the mean squared displacement. As can be inferred from Fig.~\ref{Fig:C_V_t} for the two temperatures $T=0.06$ and $T=0.30$, the microscopic time scale is $t_\mathrm{mic} \approx 0.2$. Note that the temperature dependence of $t_\mathrm{mic}$ is very weak for $T \in [0.06 \,, 0.30]$, as expected.

\begin{figure}
\includegraphics{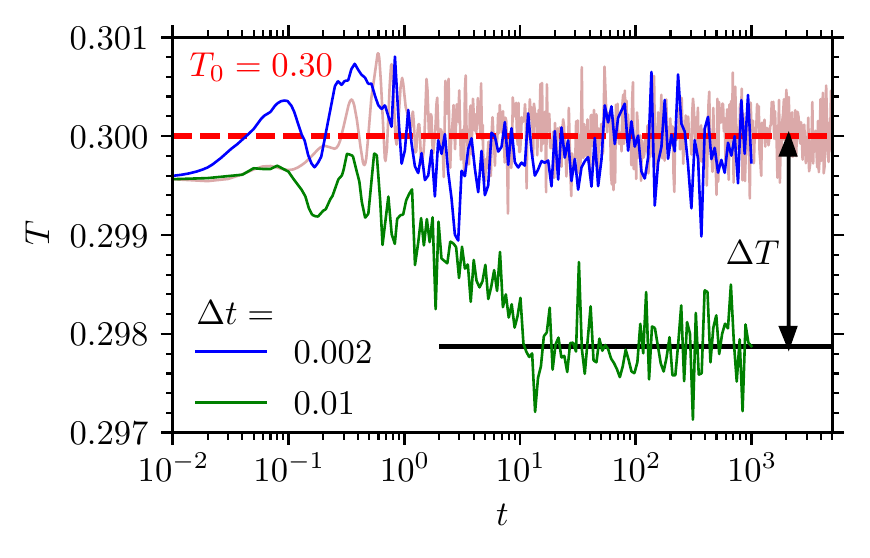}
\caption{Instantaneous temperature $T$ as a function of time $t$ at $T_\mathrm{SMC} = T_0 = 0.30$. For $NVE$ dynamics (brown curve), MD-SMC with the default time step $\Delta t = 0.01$ (green curve), and MD-SMC with $\Delta t = 0.002$ (blue curve). For the MD-SMC dynamics, $f_{\rm SMC}=1000$ with $t_\mathrm{MD} = 0.01$ is chosen. Curves are averaged over $60$ samples (each with $N=2048$ particles). \label{Fig:SMC_T_t_temperature_shift}}
\end{figure}
\textit{Temperature shift?} As already reported in Ref.~\cite{berthier2019efficient}, there can be a problem with a slight temperature shift if one chooses $t_{\rm MD} = 0.01$, which coincides with the time step $\Delta t = 0.01$. Here, we show that this is not a principal problem of the choice of a very small time $t_{\rm MD}$. Instead, it is a numerical problem with regard to the integration of the equations of motion and can be simply solved by choosing a smaller time step. This is demonstrated in Fig.~\ref{Fig:SMC_T_t_temperature_shift}, plotting the instantaneous (but sample-averaged) temperature $T$ as a function of time $t$ for the example $T_\mathrm{SMC} = T_0 = 0.30$, choosing $f_{\rm SMC}=1000$ with $t_\mathrm{MD} = 0.01$ and the two integration time steps $\Delta t = 0.01$ (green curve) and $\Delta t = 0.002$ (blue curve). For the pure $NVE$ dynamics (brown curve), the time step $\Delta t = 0.01$ is sufficiently small to maintain the correct temperature. In contrast, for the MD-SMC dynamics with the same time step, a relative temperature shift $|T - T_0|/T_0 \approx 0.7 \%$ occurs. With a smaller time step $\Delta t = 0.002$, one avoids this shift within the accuracy of our measurement. We have checked that the small shift for $\Delta t = 0.01$ has a negligible effect on the properties reported below and thus we keep using the time step $\Delta t = 0.01$ in the following.

\textit{Microscopic equilibrium.} We have seen that MD-SMC guarantees a correct thermostatting of the system, provided that the time step for the integration of the equations of motion is sufficiently small. Now we show that, after the application of swap moves in frozen coordinates, the particle velocities remain in equilibrium (during the subsequent MD time -- remember that SMC \textit{itself} does not affect the velocity distribution at all). To this end, we reconsider the velocity autocorrelation function in Fig.~\ref{Fig:C_V_t}. Here, the solid lines refer to standard $NVE$ dynamics, while the circles correspond to $NVE$ dynamics with an imposed singular distortion of the system at time $t=0$ by performing $10^3 \times N$ swap trials. That the circles are on top of the solid lines indicates that the Maxwell-Boltzmann velocity distribution is stationary during MD-SMC simulation. In this sense, MD-SMC seems to preserve microscopic equilibrium.

\subsection{Structural relaxation\label{sec:Structural_Relaxation}}
In this section we investigate the structural relaxation with MD-SMC and aim at elucidating the mechanisms how MD-SMC affects dynamic processes. The starting point for all simulations discussed below are configurations that were fully equilibrated via MD-SMC. For the analysis of the dynamics, we consider the mean squared displacement (MSD),
\begin{equation}
    \mathrm{MSD}(t) = \left \langle ( {\bf r}(t) - {\bf r}(0) )^2 \right \rangle,
    \label{eq_msd}
\end{equation}
and the self-part of the overlap function, 
\begin{equation}
     Q(t) = \left \langle \Theta( a - |{\bf r}(t) - {\bf r}(0)| ) \right \rangle.
     \label{eq_qoft}
\end{equation}
In these definitions, the angular brackets $\langle \,.\, \rangle$ indicate the particle as well as ensemble average, ${\bf r}(t)$ is the particle position vector at time $t$, $\Theta$ the Heaviside-step function, and $a=0.3$ a microscopic length scale. We use the overlap function $Q(t)$ to define a relaxation time $\tau$ via
\begin{align}
    Q(\tau) \equiv e^{-1}. 
    \label{eq:tau}
\end{align}
For the results below, we have chosen $t_\mathrm{MD} = \Delta t \equiv 0.01$ and thus we vary $f_{\rm SMC}$ via the parameter $s$, cf.~Eq.~(\ref{eq:f_SMC_density}).

\begin{figure}
\includegraphics{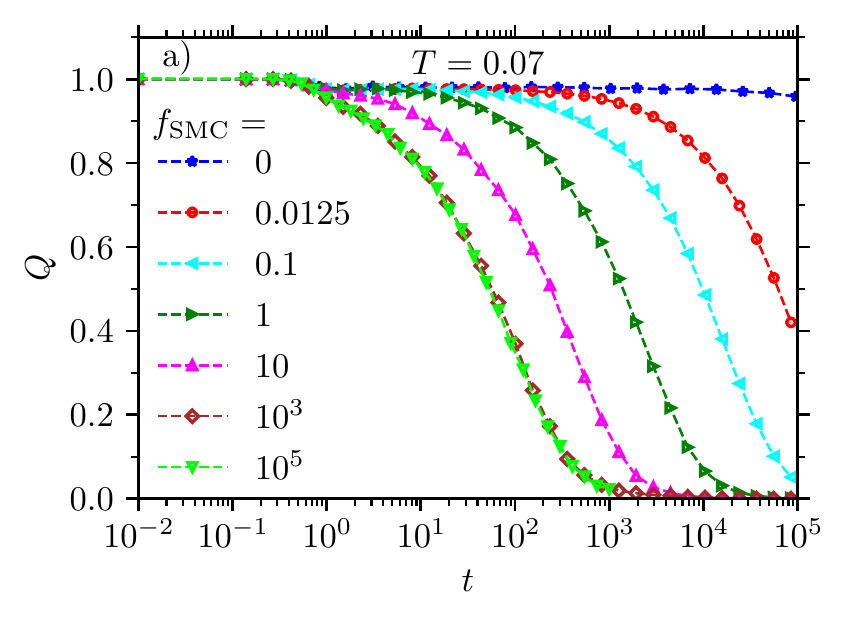}\\
\includegraphics{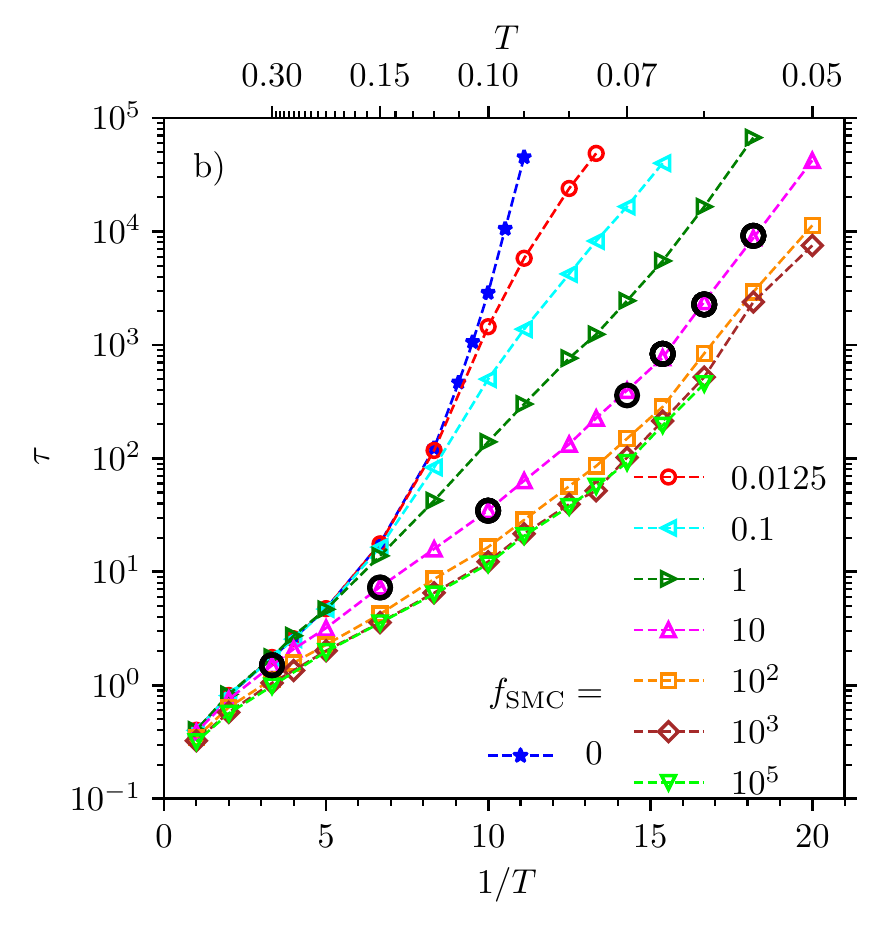}
\caption{a) Overlap function $Q$ as a function of time $t$ at the temperature $T=0.07$ for different values of $f_{\rm SMC}$. b) Relaxation time $\tau$ as a function of inverse temperature $1/T$ for the same values of $f_\mathrm{SMC}$ as in a). Systems with $N = 8000$ particles are chosen for $f_\mathrm{SMC} = 0$, $0.0125$, and $0.1$, with $N=500$ particles for $f_\mathrm{SMC} = 1$ and $10$, and with $N=256$ particles for $f_\mathrm{SMC} = 10^2$, $10^3$, and $10^5$. The black circles show results for $f_{\rm SMC} = 10$ with the time step $\Delta t = 0.002$ (otherwise the time step $\Delta t = 0.01$ is used). \label{Fig:tau}}
\end{figure}
In Fig.~\ref{Fig:tau}a, we show the time dependence of the overlap function $Q$ at a low temperature $T=0.07$. For pure $NVE$ dynamics ($f_\mathrm{SMC}=0$), we observe that $Q(t)$ falls onto a plateau the value of which is close to one. Thus the system behaves like an amorphous solid on the considered time scales. In fact, the glass-transition temperature of $NVE$ dynamics is $T_\mathrm{g}^{NVE} \approx 0.10$ if one considers time scales up to about $10^5$. Upon increasing the swap frequency $f_\mathrm{SMC}$, the time scale on which $Q(t)$ decays first rapidly decreases and eventually saturates for $f_{\rm SMC} \gtrsim 10^3$.

In Fig.~\ref{Fig:tau}b, the relaxation time $\tau$, as extracted from $Q(t)$, is displayed as a function of inverse temperature $1/T$ for different values of $f_{\rm SMC}$. Note that a similar plot is shown in Ref.~\cite{berthier2019efficient}. As pointed out in this work, even at very small frequencies (the smallest one here is $f_\mathrm{SMC} = 0.0125$, red curve) the gap in $\tau$ between $NVE$ and SMC, $\Delta \tau := \tau_{NVE} / \tau_\mathrm{SMC}$, covers several orders of magnitude at low $T$. This gap increases upon decreasing $T$. Upon increasing $f_\mathrm{SMC} \gtrsim 10^3$ (corresponding to $s \gtrsim 10$ here), there is the aforementioned saturation of $\tau$. For the most efficient parameters, we have $\Delta \tau \approx 0.5\times 10^4$ at $T=0.09$. If one extrapolates $\tau$ for $NVE$ dynamics below $T=0.09$, as done in Ref.~\cite{berthier2019efficient}, the gap $\Delta \tau$ covers many more orders of magnitude. In this sense, the gap between simulations and experiments of glassforming liquids is eventually closed. 

\textit{Asymptotic $\mathrm{SMC}_\infty$.} As observed in Fig.~\ref{Fig:tau}, variation of the number of sweeps $s \propto f_\mathrm{SMC}$, cf.~Eq.~(\ref{eq:f_SMC_density}), interpolates between two limiting cases of MD-SMC dynamics: (i) for $s=0$ (or $f_\mathrm{SMC} = 0$), pure $NVE$ dynamics is recovered and (ii) for $s\to \infty$ ($f_\mathrm{SMC} \to \infty$), MD-SMC is ``physically most efficient'' in the sense of a minimum relaxation time $\tau$. Formally, we define
\begin{align}
    \mathrm{SMC}_\infty 
    &:= 
     \mathrm{SMC}(\,t_\mathrm{MD} \to 0,~s \to \infty\,)
    \label{eq:SMC_infty_1} \\
    &\,\approx
    \mathrm{SMC}(\,t_\mathrm{MD} = 0.01,~s=10^3\,).
    \label{eq:SMC_infty_2}
\end{align}
Of course, the value $t_\mathrm{MD} = 0.01 \equiv \Delta t$ is the smallest possible value for MD simulations with a time step $\Delta t$ (with the caveat that there might be a small temperature shift if $\Delta t$ is too large, see Fig.~\ref{Fig:SMC_T_t_temperature_shift}). 

The observation that the asymptotic behavior occurs at $s \gtrsim 10 \approx s^\mathrm{rel}$ is perfectly reasonable with regard to Sec.~\ref{Sec:numerical_results_swap_on_a_grid}, where we saw that for the standard SMC the diameter autocorrelation function decays onto a plateau after a relaxation time $s^\mathrm{rel} \approx 10$, almost independently of the temperature $T$. Thus for $s \gg s^\mathrm{rel}$ we expect a ``full thermalization`` of the diameters and asymptotic behavior of MD-SMC. 
 
Note that the $\mathrm{SMC}_\infty$ is by far not the \textit{computationally} most efficient parameter setting. 

\begin{figure}	
\includegraphics{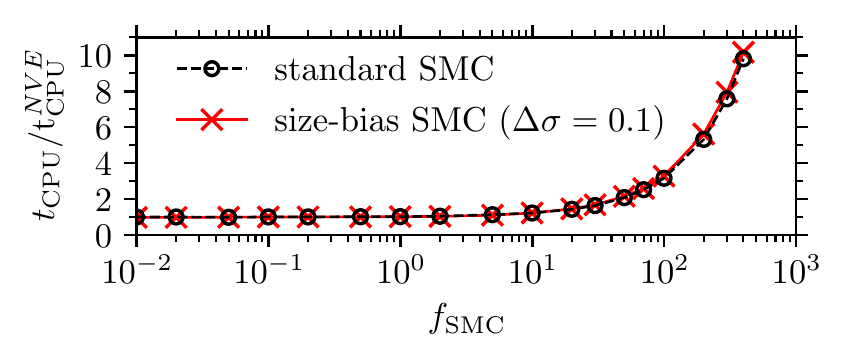}
\caption{Normalized CPU time $t_\mathrm{CPU}/t_{\rm CPU}^{NVE}$ of MD-SMC as a function of $f_\mathrm{SMC}$. Temperature $T=0.07$, $N=2048$ particles. \label{Fig:CPU_time}}
\end{figure}
\textit{Computationally most efficient SMC.} To find the frequency $f_\mathrm{SMC}^*$ for the \textit{computationally most efficient} SMC, one has to take into account the required CPU time $t_{\rm CPU}$ of MD-SMC simulations. In Fig.~\ref{Fig:CPU_time}, inspired by Ref.~\cite{berthier2019efficient}, we show $t_\mathrm{CPU}$, normalized by that of the pure $NVE$ dynamics, as a function of $f_{\rm SMC}$. Obviously, the additional computational load of the size-bias SMC compared to the standard SMC is negligible. To estimate $f_\mathrm{SMC}^*$, a reasonable approach is to minimize the product of CPU time $t_\mathrm{CPU}$ with relaxation time $\tau$. This method was proposed in Ref.~\cite{berthier2019efficient} and the authors found $f_\mathrm{SMC}^* \in [20,100]$. In the latter interval, the SMC part requires between $50\%$ and $240\%$ of the CPU time of the MD part. We can understand this range of values for $f_\mathrm{SMC}^*$ \textit{a priori} from the discussion in Sec.~\ref{Sec:numerical_results_swap_on_a_grid}, from which we expect $f_\mathrm{SMC}^* := s^\mathrm{rel}/t_\mathrm{mic} \approx 50$. A similar estimate can be obtained by $f_\mathrm{SMC}^* := s^\mathrm{rel}/t_\mathrm{vib}$, where $t_\mathrm{vib}$ is the microscopic time scale on which a particle rattles inside its cage in an amorphous solid state. The latter time scale can be estimated via $t_\mathrm{vib} := l/v_\mathrm{thm}$, with $l(T)$ a temperature-dependent localization length and $v_\mathrm{thm}(T)$ the average thermal velocity of a particle at temperature $T$. The localization length $l$ can be calculated from the MSD, see Eq.~(\ref{eq:von_schweidler}) below. For an amorphous solid at $T=0.07$, we obtain $l(T) \approx 0.063$, $v_\mathrm{thm}(T) \approx \sqrt{k_\mathrm{B}T/m} \approx 0.26$, and thus $f_\mathrm{SMC}^* = s^\mathrm{rel} v_\mathrm{thm}/l \approx 40$.

\begin{figure}	
\includegraphics{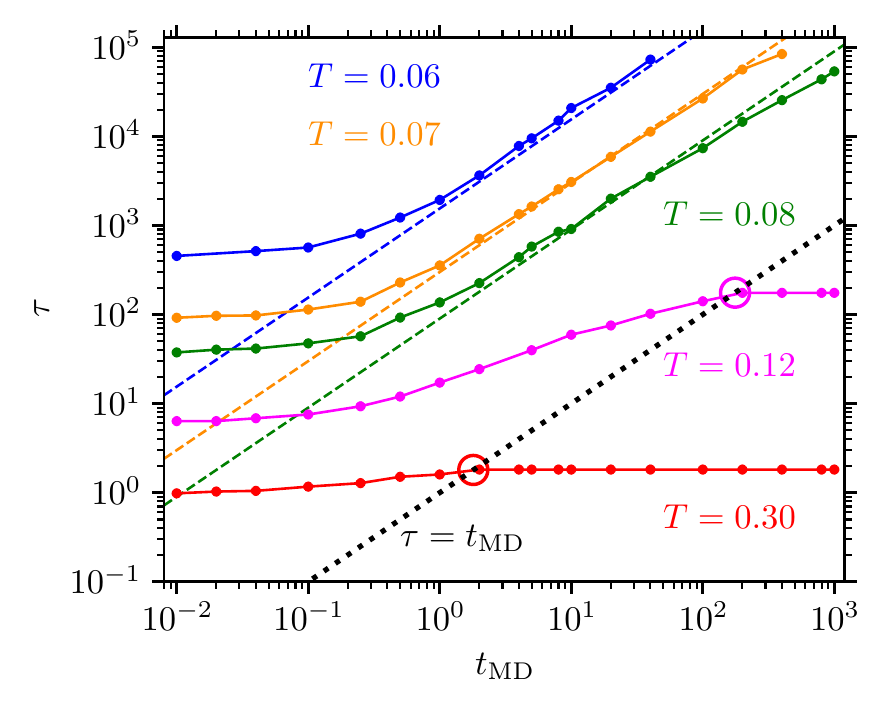}
\caption{Relaxation time $\tau$ as a function of $t_\mathrm{MD}$ for different temperatures $T$. A fixed value $s=10^3$ and $N=2048$ particles are used. The dashed lines indicate a proportionality $\tau \propto t_\mathrm{MD}$, while the dotted black line shows $\tau = t_\mathrm{MD}$. \label{Fig:tau_tMD}}
\end{figure}
\textit{Variation of $t_\mathrm{MD}$.} Now we analyze how the MD-SMC dynamics changes under the variation of $t_\mathrm{MD}$, keeping the number of swap sweeps fixed to large value $s=10^3$. As we shall see below, the choice $t_\mathrm{MD} \gg t_\mathrm{mic} \approx 0.2$ enables us to disentangle the Newtonian dynamics of MD from the effect of swapping with SMC.

Figure~\ref{Fig:tau_tMD} shows the relaxation time $\tau$ as a function of $t_\mathrm{MD}$ for different temperatures $T$. For all considered temperatures, the relaxation time increases upon increasing $t_\mathrm{MD}$, as the diameters are thermalized less frequently, reducing the effect of SMC on the dynamics. Upon increasing $t_\mathrm{MD}$ beyond the microcanonical ($NVE$) relaxation time $\tau_{NVE}$, the chronologically first full thermalization of the diameters at $t = t_\mathrm{MD}$ has no influence on the calculation of the relaxation time $\tau$, see Eq.~(\ref{eq:tau}). Thus we have $\tau_\mathrm{SMC} \equiv \tau_{NVE}$ for $t_\mathrm{MD} \geq \tau_{NVE}$. The equality can numerically only be observed for the two higher temperatures, as for the other $T$ the $NVE$ relaxation times are beyond the viable simulation time.

Now let us analyze the other limit: upon decreasing $t_\mathrm{MD}$ and approaching a microscopic time scale, $t_\mathrm{MD} \approx t_\mathrm{mic} \approx 0.2$,  a saturation sets in. For $t_\mathrm{MD} = 0.01 \equiv \Delta t$ the curves have converged within numerical precision. In the previous subsection we showed that, for any given $T$, $\mathrm{SMC}(t_\mathrm{MD}=0.01,\,s)$ has numerically converged with respect to $s$ if $s \gtrsim 10^3$. Now we see that $\mathrm{SMC}(t_\mathrm{MD},\,s=10^3)$ has also numerically converged with respect to the parameter $t_\mathrm{MD}$ when close to $0.01$. We can conclude that $\mathrm{SMC}(t_\mathrm{MD} = 0.01,\,s=10^3)$ in fact resembles the converged $\mathrm{SMC}_\infty$ dynamics up to a decent numerical precision, confirming Eq.~(\ref{eq:SMC_infty_2}).

A remarkable observation in Fig.~\ref{Fig:tau_tMD} is that a linear regime develops for $T$ below the microcanonical glass-transition temperature $T^{NVE}_\mathrm{g} \approx 0.10$. It seems that $\tau \propto t_\mathrm{MD}$ when $t_\mathrm{mic} < t_\mathrm{MD} < \tau_{NVE}$. To understand this observation, we analyze the $\mathrm{MSD}$ at a low temperature $T=0.07$ in the next subsection.

\subsection{The relaxation mechanism of MD-SMC\\in an amorphous solid}
We saw that hybrid MD-SMC is particularly efficient at low temperatures, i.e.~at temperatures $T$ far below the glass-transition temperature $T^{NVE}_\mathrm{g} \approx 0.10$ of a conventional MD simulation. This is possible since MD-SMC opens a new relaxation channel, the origin of which shall be revealed in the following. For this purpose we consider the temperature $T=0.07$, which was characterized as an amorphous solid state of pure MD before: in Fig.~\ref{Fig:tau}a, we showed that the overlap function $Q(t)$ has a pronounced plateau-like region up to a time scale $t \approx 10^4$.

\begin{figure}
\includegraphics{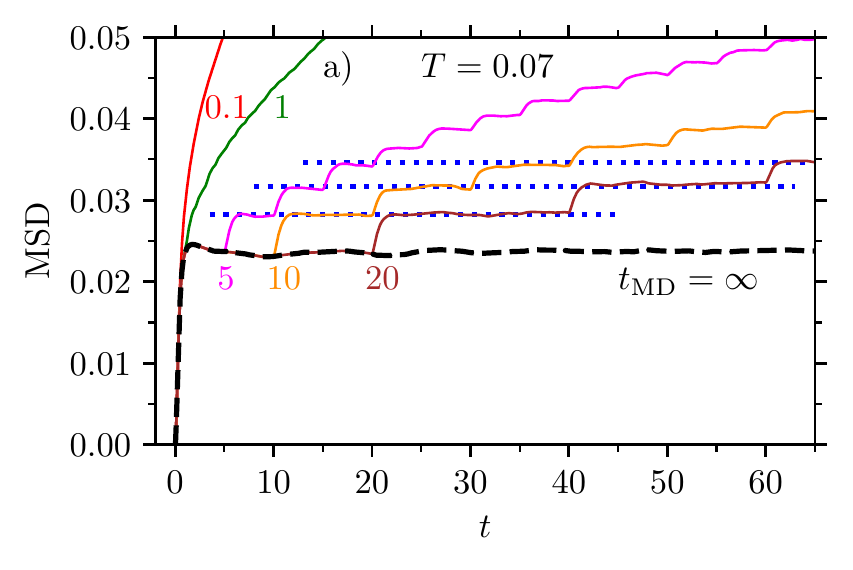}\\
\includegraphics{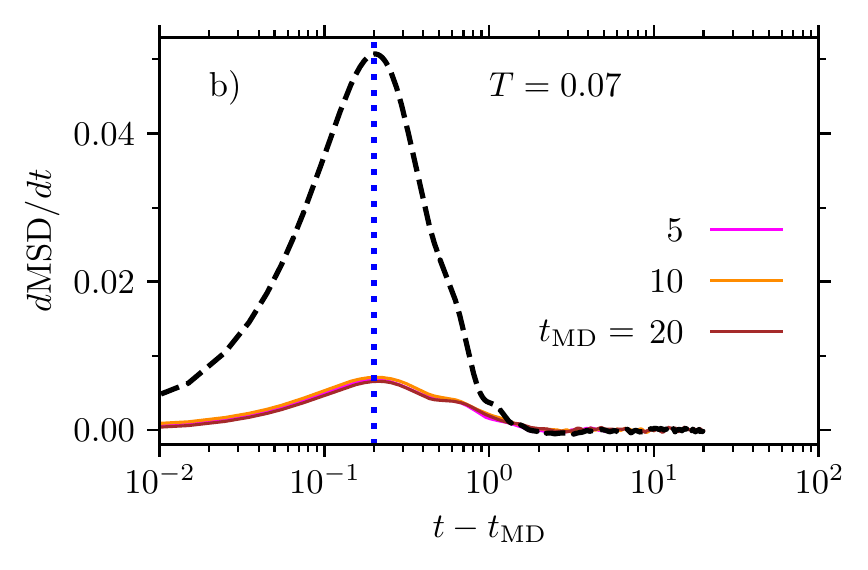}
\caption{a) MSD as a function of time $t$ for different values of $t_\mathrm{MD}$, as indicated by the colored numbers. Blue dotted lines act as a guide to the eye. b) The derivative $d\mathrm{MSD}/dt$ as a function of $t - t_\mathrm{MD}$ for finite values of $t_\mathrm{MD}$ and as a function of $t$ for $t_\mathrm{MD}=\infty$ (black dashed line). The vertical dotted line marks the microscopic time $t_{\rm mic}=0.2$. In both panels, number of swap sweeps $s=10^3$ and $N=8000$ particles. \label{fig:MSD_t_3}}
\end{figure}

In Fig.~\ref{fig:MSD_t_3}a, we show the mean squared displacement (MSD) as a function of time $t$ for relatively short times, $t<10^2$. For $NVE$ dynamics (black dashed curve, $t_\mathrm{MD} = \infty$), the MSD exhibits a plateau, which quantifies the localization of each particle inside its cage. The colored curves and numbers represent the hybrid MD-SMC dynamics for varying $t_\mathrm{MD}$. After every $t_\mathrm{MD}$, we perform a full thermalization of the diameters, i.e., $10^3 \times N$ swap moves are attempted. An intriguing feature of the MSDs in Fig.~\ref{fig:MSD_t_3}a is that, after every $t_\mathrm{MD}$, there is a jump of the plateau value to a higher level (for $t_\mathrm{MD} \geq 5$). Here the MSD at time $t \in [ n t_\mathrm{MD},\,(n+1) t_\mathrm{MD} ]$ only depends on $n \in \mathbb{N}$, the number of jumps or diameter thermalizations, as indicated by the blue horizontal lines. This explains the linear regime $\tau \propto t_\mathrm{MD}$ observed in Fig.~\ref{Fig:tau_tMD} and Ref.~\cite{berthier2019efficient}. The time scale $t_\mathrm{jmp}$ of the jumps, i.e.~the relaxation time from one plateau to the next, is short but finite.

In Figure~\ref{fig:MSD_t_3}b, we quantify the time scale $t_\mathrm{jmp}$ of the jumps, by plotting the derivative $d\mathrm{MSD}/dt$ as a function of $t - t_\mathrm{MD}$ for different values of $t_\mathrm{MD}$. For $t_\mathrm{MD} = \infty$ (black dashed line), we show $t$ on the $x$-axis instead. We can infer from the figure, that the time scale $t_\mathrm{jmp}$ coincides with the microscopic time scale $t_\mathrm{mic}\approx 0.2$ (vertical blue line) on which a particle relaxes within its cage as a consequence of collisions with its neighbors during MD.

For $t_\mathrm{MD} \lesssim 1$, the time scale $t_\mathrm{MD}$ starts to interfere with the microscopic time scale $t_\mathrm{mic}\approx 0.2$. Here, $t_\mathrm{MD}$ is too short to allow a complete relaxation onto a new plateau before just another thermalization of the diameters is imposed by SMC. Thus, in Fig.~\ref{fig:MSD_t_3}a, the phenomenology of a stepwise relaxation vanishes for $t_\mathrm{MD} \lesssim 1$.

\begin{figure}
\centering
\begin{minipage}{.5\columnwidth}
\centering
\includegraphics[width=.95\columnwidth]{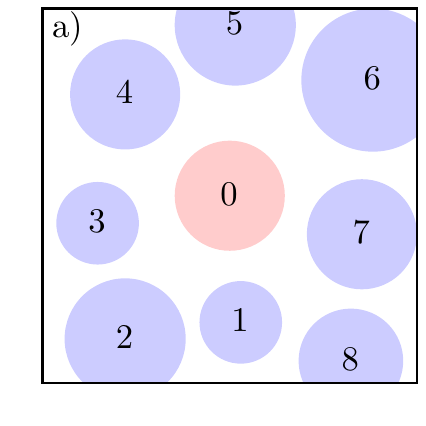}
\end{minipage}%
\begin{minipage}{.5\columnwidth}
\centering
\includegraphics[width=.95\columnwidth]{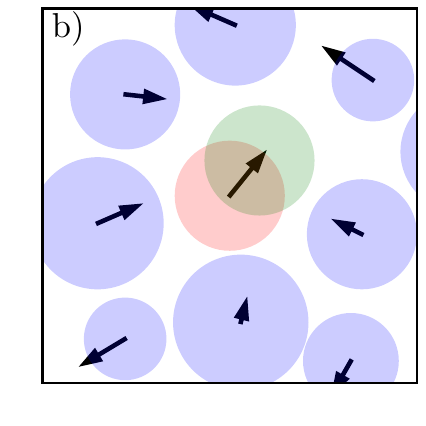}
\end{minipage}
\caption{Schematic illustration of the proposed relaxation mechanism of the MD-SMC dynamics. A sketch of a two-dimensional system is shown a) just before SMC, b) immediately after SMC; the particle positions are exactly the same in a) and b). Before SMC, each particle is shown at its assumed time-averaged position. After SMC, each particle finds itself in a new cage geometry, and thus each mean position has changed. The corresponding shifts are indicated by black arrows. As an example, for the red sphere labeled by $0$, its new mean position is indicated by the green sphere. \label{Fig:mechanism_scheme}}
\end{figure}
\textit{Relaxation mechanism.} The stepwise increase of the MSD reveals the origin of the very efficient structural relaxation at low temperatures via the MD-SMC dynamics. The occurrence of a plateau in the MSD indicates a ``frozen'' structure where each particle is localized in a cage formed by surrounding neighboring particles. A step in the MSD after a thermalization with swap moves is associated with a rearrangement of the cage structure: As shown in Sec.~\ref{Sec:theory_swap_on_a_grid}, the application of SMC on a configuration with fixed particle coordinates assigns a new \textit{equilibrium} permutation of diameters and thereby a new cage geometry around each particle is imposed. After SMC, during MD over the time $t_\mathrm{MD}$, each particle continues to perform vibrations in a cage, however, and this is the crucial point, \textit{now within a differently shaped cage}. Here the particle can explore a (slightly) different region in coordinate space. The relaxation toward a new mean position, as manifested by a jump in the MSD, occurs on a microscopic time scale $t_\mathrm{mic}$, cf.~Fig.~\ref{fig:MSD_t_3}. The proposed mechanism clarifies the drastic speed-up of the dynamics: While the diameters are exchanged \textit{instantaneously} during the SMC part, the subsequent relaxation within a new cage occurs on a short \textit{microscopic time scale}. 

To reveal this relaxation mechanism, we disentangled the SMC from the MD part by choosing a relatively large (computationally inefficient) value $t_\mathrm{MD} > t_\mathrm{mic} \approx 0.2$. The stepwise increase of the MSD turns into a continuous increase for small values of $t_\mathrm{MD}$, cf.~Fig.~\ref{fig:MSD_t_3}a. Here MD-SMC is most efficient.

Figure \ref{Fig:mechanism_scheme} schematically illustrates the MD-SMC relaxation mechanism. Before SMC, a), each particle is trapped inside a cage with a specific geometry. We show each particle at its assumed time-averaged position. In b), after the diameters were swapped via SMC, the cage geometry around each particle has changed. Thus, during the subsequent MD part, each particle will fluctuate around a new mean position. The corresponding shifts of the mean positions are indicated by black arrows. As an example, for the red sphere labeled by $0$, the green sphere shows its new average position after SMC. In the illustration, we purposely did not change the diameter of particle $0$. Thereby we want to emphasize that the altered cage environment of a tagged particle is the essential ingredient to the relaxation mechanism. In this sense the mechanism is consistent with the finding in Ref.~\cite{ninarello2017}, that the displacement of a tagged particle via SMC is not always linked to a change of its own diameter.

Above, we identified the time scale $t_\mathrm{jmp} = t_\mathrm{mic} \approx 0.2$ on which the jumps in the MSD occur. To quantify the distribution of jump lengths, we measure shifts $\Delta \bar{x}$ of subsequent mean positions $\bar{x}_n$, which are triggered by the application of SMC. Here, $\Delta \bar{x} = \bar{x}_{n+1}- \bar{x}_{n}$, where $\bar{x}_n = t_\mathrm{MD}^{-1} \int_{nt_\mathrm{MD}}^{(n+1)t_\mathrm{MD}} x(t)\,dt$ is calculated by averaging the $x$-coordinate of a particle over the $n$-th MD block of time span $t_\mathrm{MD}$. These definitions are robust when $t_\mathrm{mic} \ll t_\mathrm{MD} \ll \tau_{NVE}$.

\begin{figure}
\centering
\includegraphics{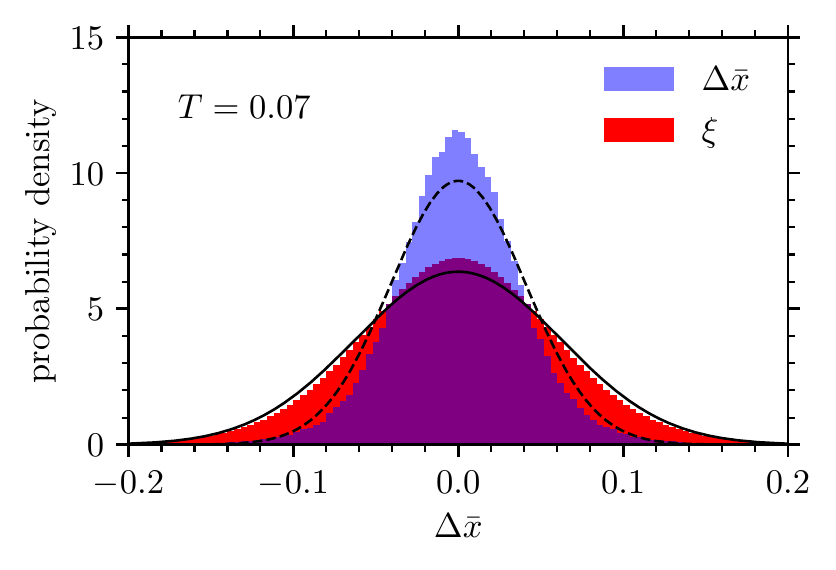}
\caption{Probability distributions of $\Delta \bar{x} = \bar{x}_{n+1}- \bar{x}_{n}$ (blue) and of $\xi(t) = x(t) - \bar{x}_n$ (red). For both quantities, a fit of a normal distribution is shown (black lines). Results are calculated from a sample with $N=8000$ particles, considering $10$ MD time-blocks, each of length $t_\mathrm{MD} = 100$. \label{fig:zeta_histogram}}
\end{figure}

In Fig.~\ref{fig:zeta_histogram}, we show the distribution of $\Delta \bar{x}$ (blue), considering many particles and MD blocks. As a reference, we show a zero-centered normal distribution (black dashed line) with a variance calculated from the data.

We want to compare $\Delta \bar{x}$ with the fluctuations of the particles inside their cage during the MD part. To this end, we introduce $\xi(t) = x(t) - \bar{x}_n$ on the interval $t \in [nt_{\rm MD},(n+1)t_{\rm MD}]$. The distribution of $\xi$ over many particles and MD blocks is shown in Fig.~\ref{fig:zeta_histogram} (red). Also here, we plot a zero-centered normal distribution (black solid line) with a variance calculated from the data. We find that the distribution of the mean-position shifts $\Delta \bar{x}$ is comparable to, but slightly narrower than, the fluctuations $\xi$ of the particles inside their cage. Note that the variance of $\xi$ (with $\langle \xi \rangle = 0$) is related to the MSD via $\langle \xi^2 \rangle = \frac{1}{6} \mathrm{MSD} = l^2$. Here we introduced a localization length $l$, which is in the focus of the next section.

%
\subsection{Glass transition: $NVE$ vs.~$\mathrm{SMC}_\infty$}
We saw that the stepwise increase of the MSD is associated with a sequence of rearrangements of the cage structure around each particle. In each of these steps, the particles shift to new mean positions. However, this dynamic process cannot be described as a random walk, since the new configuration after the rearrangement of cages is still strongly correlated with the previous one. This correlation manifests in a shoulder of the $\mathrm{MSD}$ on intermediate time scales -- even when the physically most efficient SMC [$\mathrm{SMC}_\infty$, see Eq.~(\ref{eq:SMC_infty_1})], is used. At sufficiently low $T$, we can identify a plateau in the MSD also for $\mathrm{SMC}_\infty$ dynamics. In this section, we extract the associated length scale with a \textit{von Schweidler law} and compare it with that of $NVE$ dynamics.

Note that now we include temperatures $T < T_\mathrm{g}^{\rm SMC} \approx 0.06$ below the glass-transition temperature. Here, our preparation protocol does not provide fully equilibrated configurations anymore.

\begin{figure}
\includegraphics{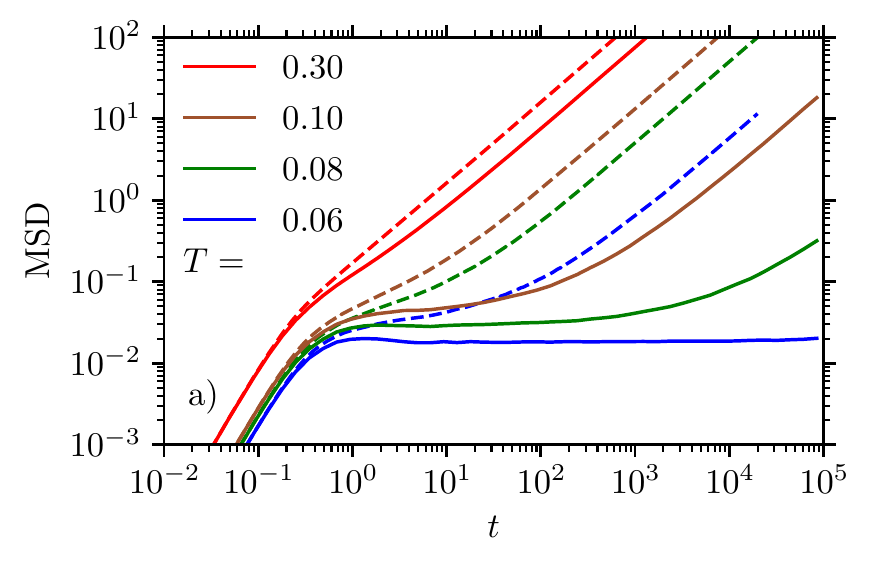}
\includegraphics{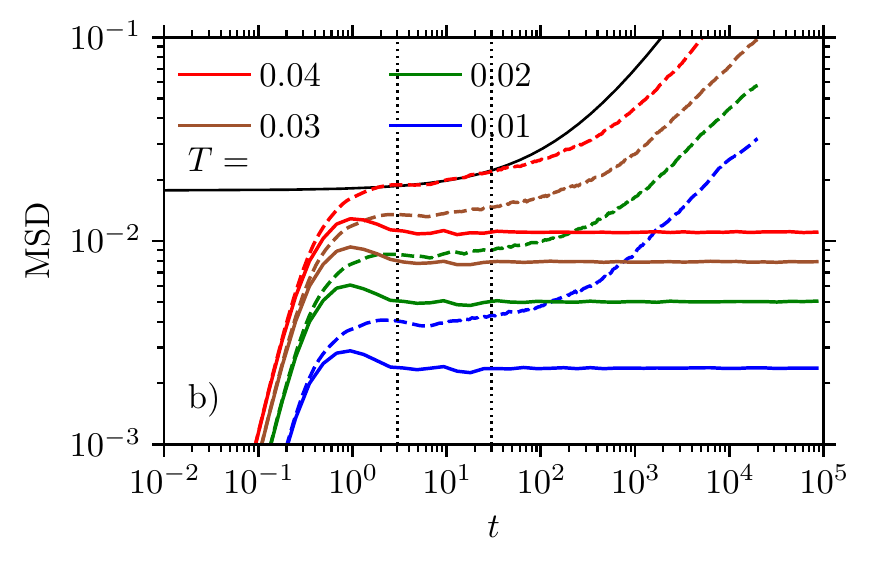}
\caption{Mean squared displacement $\mathrm{MSD}$ as a function of time $t$ for $NVE$ (solid lines) and  $\mathrm{SMC}_\infty$ dynamics (dashed lines) for different temperatures $T$, respectively. In a) equilibrium data (higher $T$), while in b) \textit{non}-equilibrium curves (lower $T$) are shown. For $T=0.04$ and $\mathrm{SMC}_\infty$, we fit a solid black curve according to \textit{von Schweidler}, cf.~Eq.~(\ref{eq:von_schweidler}). The fitting interval is indicated by dotted vertical lines. \label{Fig:MSD_t}}
\end{figure}

In Fig.~\ref{Fig:MSD_t}, we show the MSD as a function of time $t$ for $NVE$ (solid lines) and $\mathrm{SMC}_\infty$ dynamics (dashed lines). We cover a wide range of temperatures: In Fig.~\ref{Fig:MSD_t}a, results for four higher temperatures $T$ are displayed, for which the initial configurations were fully equilibrated with MD-SMC. In Fig.~\ref{Fig:MSD_t}b, the dynamics for lower temperatures are shown, where the initial configurations could not be fully equilibrated. We observe the typical phenomenology of glassforming liquids, i.e., the MSDs develop shoulders and plateaus upon decreasing the temperature $T$. These plateaus reflect the localization of the particles in a cage. Their height describes a characteristic squared length scale which is significantly smaller than the squared nearest-neighbor distance between particles. 

Interestingly, in Fig.~\ref{Fig:MSD_t}b it seems that the overshoots in the MSDs around a time $t \approx 1$ for $NVE$ dynamics are absent for $\mathrm{SMC}_\infty$. Since the overshoot is associated with particle vibrations, its absence in MD-SMC seems plausible in consideration of the proposed relaxation mechanism, which qualitatively changes the cage dynamics.

%
\textit{Localization length $l$.} Mode coupling theory (MCT) \cite{gotze2009complex} predicts the asymptotic behavior of the MSD around the plateau region. According to this theory, the initial increase from the plateau toward the diffusive regime is given by a von Schweidler law. This is a power law that can be seen as a fingerprint of glassy dynamics. It reads
\begin{equation}
    \mathrm{MSD}(t) = 6 l^2 + c t^b, 
    \label{eq:von_schweidler}
\end{equation}
where the exponent $b$ is predicted to be universal for a given system and $c>0$ is a critical amplitude. We use fits to Eq.~(\ref{eq:von_schweidler}) to estimate the temperature dependence of the localization length $l$ from the MSDs. We choose $b=0.7$ and a time interval $t \in [3, 30]$. Then the parameters $l^2$ and $c$ appear linear in Eq.~(\ref{eq:von_schweidler}) and thus they can be calculated via a linear regression model. An example of such a fit for the temperature $T=0.04$ and $\mathrm{SMC}_\infty$ dynamics is shown in Fig.~\ref{Fig:MSD_t}b. 

\begin{figure}
\includegraphics{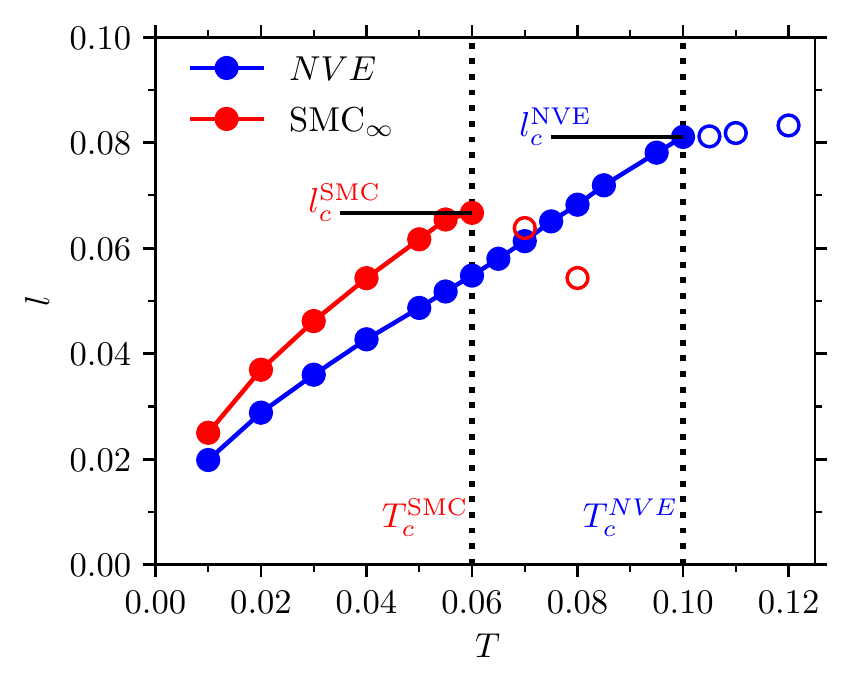}
\caption{Localization length $l$ as a function of temperature $T$ for $NVE$ and $\mathrm{SMC}_\infty$ dynamics. The plateau height $l$ is calculated from the MSDs via a \textit{von Schweidler} fit, see Fig.~\ref{Fig:MSD_t}. The vertical lines indicate the critical temperatures $T_\mathrm{c}$. \label{Fig:localization_length_T}}
\end{figure}

Figure \ref{Fig:localization_length_T} shows the localization length $l$ as a function of temperature $T$ for $NVE$ dynamics (blue circles) and $\mathrm{SMC}_\infty$ (red circles). The vertical lines show the respective critical MCT temperatures $T_\mathrm{c}$, identified by the maximal $T$ beyond which the fitting procedure (subjectively) fails. MCT predicts that coming from temperatures $T<T_c$ the localization length reaches a critical value $l_c$ at $T_c$ that marks the stability limit of the amorphous solid state and thus above $T_c$ the system is in a liquid state.

In experiments and simulations of glassforming liquids, an ideal glass transition, as predicted by MCT, does not exist. In real systems, the temperature $T_c$ can be interpreted as a crossover temperature between a liquid-like dynamics for $T>T_c$ and a solid-like dynamics for $T<T_c$. We use the von Schweidler law (\ref{eq:von_schweidler}) with the aim to determine the localization length $l$. While at low temperatures the estimated values for $l$ are very robust, the fitting procedure becomes more problematic at higher temperatures where a plateau or even a shoulder in the MSD can hardly be identified. However, this behavior of the MSD manifests the gradual crossover from a solid- to a liquid-like dynamics with increasing temperature. In the real system, when the temperature is increased from below $T_c$, the stability limit is associated with a vanishing life time of the amorphous solid state and thus the plateau in the MSD gradually disappears when $T_c$ is approached \cite{lamp2022}.   

In correspondence with MCT, we observe a saturation of the localization length at the critical values $l_c^{\rm SMC} \approx 0.067$ for $\mathrm{SMC}_\infty$ and $l_c^{NVE}\approx 0.081$ for $NVE$ dynamics. That the critical localization length $l_c$ is significantly larger for the $\mathrm{SMC}_\infty$ than for the $NVE$ dynamics is in agreement with the theoretical prediction of Szamel \cite{szamel2019theory} in the framework of an MCT model. 
 
We find $l^{\rm SMC} > l^{NVE}$ (for $T < T_c^{\rm SMC}$, where the comparison is meaningful). This result can be understood with a simple geometric picture: Let us pin all coordinates of all particles except for one tagged particle. When the diameters of the particles which form a cage around the tagged one fluctuate, the tagged particle can explore a slightly larger region in its cage than without SMC.

We can also infer from Fig.~\ref{Fig:localization_length_T} that toward low temperatures the localization lengths of $\mathrm{SMC}_\infty$ and $NVE$ dynamics tend to approach each other. In fact, this is expected from the geometric picture above and our finding in Fig.~\ref{Fig:pure_SMC_C_plateau_T} that the plateau value of $C_\sigma$ approaches $1$ for $T \to 0$. This explains that the thermalization of diameters has a diminishing effect on the localization length $l$ with decreasing temperature.

\section{Conclusions \label{sec:conclusions}}
In this work, we have investigated a polydisperse model glassformer by augmenting MD simulations with SMC. Our aim has been to reveal the mechanisms by which MD-SMC allows to obtain equilibrated states at very low temperatures that are far below the glass-transition temperature of any viable pure MD. In fact, ultrastable states can be generated that are comparable to those realized in typical experiments of glassforming systems. As we have shown in this work, this is possible because the MD-SMC dynamics qualitatively changes the caging of each particle in a dynamic manner while it provides a proper equilibrium sampling. As a consequence, the glass transition as identified via the critical MCT temperature $T_c$ shifts to a much lower temperature when compared to pure Newtonian dynamics and the critical localization length at $T_c$ is significantly lower, $l_c^{\rm SMC} \approx 0.067 < l_c^{NVE} \approx 0.081$.

A central idea of our study has been to disentangle the effect of swap moves from the exploration of coordinate space via Newtonian dynamics.  To this end, we first studied SMC on a frozen configuration. Here, we elaborated a full mathematical description of SMC as a Metropolis-Hastings algorithm on a confined phase space of particle permutations. Three different SMC variants (standard, size-bias, local) were introduced and characterized by symmetric proposal probabilities. For each variant, we discussed the conditions under which the Markov chain converges to the target distribution. Numerically we compared the performance of each SMC variant with a diameter correlation function and its relaxation time $s_{\rm rel}$. For the standard SMC, we found that $s_{\rm rel} \approx 10$ swap sweeps are required to ``thermalize'' the diameters on a frozen configuration. For the size-bias SMC, we found the optimized parameter $\Delta \sigma \approx 0.1$. At a low temperature, this optimized size-bias SMC only requires about $1/4$ of the swap trials of the standard SMC. The local SMC scheme tends to have the worst performance for the considered polydisperse system, but if one chooses $\Delta r \gtrsim 3$ for the range of the local SMC, it is as efficient as the standard SMC. The local SMC is particularly interesting as a possible candidate for a parallel implementation of SMC for large systems, with the option to optimize the efficiency by combining it with the size-bias SMC.

For the hybrid MD-SMC dynamics, we have shown that it is not necessary to use an additional thermostat (provided that the time step $\Delta t$ is sufficiently small); MD-SMC itself guarantees a proper thermostatting of the system. To implement the physically most efficient MD-SMC, denoted by $\mathrm{SMC}_\infty$ above, the time $t_{\rm MD}$ between swap sweeps is as small as possible (i.e., $t_{\rm MD}=\Delta t$) and in each swap round at least $s_{\rm rel}$ sweeps are performed.

We have shown how SMC qualitatively changes the dynamics at low temperatures by choosing $t_{\rm MD}$ such that it is significantly larger than the microscopic time scale $t_{\rm mic}\approx 0.2$. Then, the MSD shows a stepwise increase with MD-SMC (instead of a single plateau for pure MD dynamics). At each of these steps, a new diameter permutation is instantaneously imposed with SMC, changing the cage geometry around each particle. Then, during MD, a shift of the mean position of each particle occurs on the microscopic time scale $t_{\rm mic}$. It is this mechanism that explains the drastic speed-up of the dynamics.


\bibliography{biblio}

\end{document}